\documentclass[aip,reprint,amsmath,amssymb]{revtex4-1}

\usepackage{graphicx}
\usepackage{dcolumn}
\usepackage{float}
\usepackage{bm}
\usepackage{amssymb}
\usepackage{lipsum}
\usepackage[usenames,dvipsnames]{xcolor}

\begin{document}

\title{Equilibrium behaviour of two cavity-confined polymers: Effects of polymer width and 
system asymmetries}

\author{Desiree A. Rehel}
\email{reheld@mcmaster.ca}
\affiliation{ Department of Physics, University of Prince Edward Island,
550 University Avenue, Charlottetown, Prince Edward Island, C1A 4P3, Canada.}
\altaffiliation{Present address: Department of Physics \& Astronomy,
McMaster University, 1280 Main Street West, Hamilton, Ontario, L8S 4K1, Canada.}
\author{James M. Polson}
\email{jpolson@upei.ca}
\affiliation{ Department of Physics, University of Prince Edward Island,
550 University Avenue, Charlottetown, Prince Edward Island, C1A 4P3, Canada }

\begin{abstract}
Experiments using nanofluidic devices have proven effective in characterizing the physical
properties of polymers confined to small cavities.  Two recent studies using such methods
examined the organization and dynamics of two DNA molecules in box-like cavities with strong
confinement in one direction and with square and elliptical cross sections in the
lateral plane.  Motivated by these experiments, we employ Monte Carlo and Brownian dynamics
simulations to study the physical behaviour of two polymers confined to small
cavities with shapes comparable to those used in the experiments. We quantify the
effects of varying the following polymer properties and confinement dimensions on the
organization and dynamics of the polymers: the polymer width, the polymer contour length ratio,
the cavity cross-sectional area, and the degree of cavity elongation for cavities with
rectangular and elliptical cross sections. We find that the tendency for polymers to
segregate is enhanced by increasing polymer width.  For sufficiently small cavities, increasing
cavity elongation promotes segregation and localization of identical polymers to
opposite sides of the cavity along its long axis. A free-energy barrier controls the rate of
polymers swapping positions, and the observed dynamics are roughly in accord with predictions
of a simple theoretical model. Increasing the contour length difference between polymers significantly affects
their organization in the cavity. In the case of a large linear polymer co-trapped with a small
ring polymer in an elliptical cavity,  the small polymer tends to lie near the lateral confining
walls, and especially at the cavity poles for highly elongated ellipses. 
\end{abstract}

\maketitle

\section{Introduction}
\label{sec:intro}

Confinement of a polymer chain to a sufficiently small space significantly affects its
size and shape as well as its dynamical properties. The resulting conformational distortion 
tends to enhance the entropic repulsion between different polymer chains, thus reducing
interchain overlap and promoting chain segregation. A well studied case is confinement of 
multiple polymers to long cylindrical channels and cylindrical cavities of finite length.
Numerous computer simulation studies have examined the effects of entropic repulsion for
two polymers in such geometries.\cite{jun2006entropy, teraoka2004computer, jun2007confined, 
arnold2007time, jacobsen2010demixing, jung2010overlapping, jung2012ring, jung2012intrachain, 
liu2012segregation, dorier2013modelling, racko2013segregation, shin2014mixing, 
minina2014induction, minina2015entropic, chen2015polymer, polson2014polymer, du2018polymer, 
polson2018segregation, nowicki2019segregation, nowicki2019electrostatic, polson2021free, 
mitra2022polymer,mitra2022topology}
{Most of these studies have examined the segregation behaviour of
flexible linear polymers,\cite{jun2006entropy,teraoka2004computer,jun2007confined,%
arnold2007time,jacobsen2010demixing,jung2010overlapping,jung2012intrachain,%
liu2012segregation,racko2013segregation,minina2014induction,polson2014polymer,du2018polymer}
though some have also investigated the behaviour of ring polymers\cite{jung2012ring,%
dorier2013modelling,shin2014mixing,minina2015entropic,chen2015polymer,polson2018segregation}
and more complex topologies.\cite{mitra2022polymer,mitra2022topology}
In addition, the effects of bending rigidity,\cite{racko2013segregation,polson2014polymer}
a difference in the contour lengths,\cite{polson2021free}
macromolecular crowding,\cite{shin2014mixing,chen2015polymer,polson2018segregation}
and electrostatics\cite{nowicki2019segregation,nowicki2019electrostatic} on the
segregation dynamics and thermodynamics have been examined in detail.}
Information gleaned from these studies may be useful in helping advance 
nanofluidic technology used in applications such as DNA sorting,\cite{dorfman2012beyond} DNA 
denaturation mapping,\cite{reisner2010single, marie2013integrated} and genome 
mapping.\cite{lam2012genome, hastie2013rapid, dorfman2013fluid, muller2017optical} 
The degree to which polymers either mix or partition in such nanofluidic 
devices may impact their performance in cases where the devices are operated at high polymer 
concentrations to increase throughput.

Understanding the behaviour of confined polymers may also help elucidate the role of entropy 
in the segregation of replicating chromosomes of prokaryotes into the daughter cells. 
Some bacteria such as {\it E. coli} possess no known active machinery to promote segregation,
and it has been proposed that entropy provides the key driving force in such 
systems.\cite{jun2006entropy, jun2010entropy, diventura2013chromosome, youngren2014multifork, 
mannik2016role} In this scenario, confinement-enhanced entropic repulsion between chromosomes 
pushes them apart along the long axis of the rod-shaped cells in a manner comparable to
the process observed for two channel-confined polymers in various simulation studies. 
Several recent experimental studies 
have reported supporting evidence for such a mechanism.\cite{diventura2013chromosome, 
mannik2016role, cass2016escherichia, wu2019cell, wu2020geometric, elnajjar2020chromosome, japaridze2020direct, 
liang2020artificial, gogou2021mechanisms}  For example, Wu {\it et al.} \cite{wu2020geometric} 
found that segregation defects observed in cell-wall-less states of {\it Bacillus subtilis} could
be eliminated by confining the cells to narrow channels, which promoted successful segregation
by recovering the elongated cell shape. Likewise, Liang {\it et al.} studied chromosome segregation 
in \textit{E. coli} confined to microchannels of variable width and found the efficiency of 
chromosome segregation increases appreciably as the channel narrows. As simulations show
that the entropic driving force increases with decreasing channel width \cite{polson2014polymer, 
polson2018segregation, polson2021free} the experiments clearly support models of cell division 
in which entropy plays a prominent role in prokaryotic chromosome segregation.

While simulation studies of polymer segregation provide some degree of insight into experimental 
measurements of bacterial chromosome segregation, direct quantitative comparison is difficult. 
Bacterial chromosomes are highly complex structures packed into a crowded environment, and their 
detailed physical behaviour is difficult to capture using the simplistic bead-spring-type models that 
computer simulations typically employ.  Fortunately, a much cleaner experimental test of theoretical 
predictions and simulation results for such systems is possible using nanofluidic devices.
Recent advances in nanofabrication techniques
have facilitated the construction of devices that are ideal for trapping and manipulating individual
polymers. For example, nanotopography features such as `nanopits' can be embedded in one surface 
of a nanoslit and function as entropic traps for polymers confined to the slit. Such devices
can be used to manipulate polymer chains in various ways. For example, recent studies by 
Klotz {\it et al.} used fluorescence microscopy to examine the behaviour of single DNA chains 
confined to such a complex geometry and observed polymer contour sharing between multiple adjacent 
nanopits in a `tetris'-like conformation.\cite{klotz2015correlated, klotz2015measuring}
Alternatively, active nanofluidic approaches, such as the Convex-lens Induced 
Confinement\cite{berard2014convex} can be used to dynamically adjust the slit width, creating 
a top-loading effect whereby entire molecules can be driven inside the nanopits.

Recently, Reisner and coworkers employed a top-loading version of a nanofluidic 
structure using pneumatically activated nanoscale nitride membranes.\cite{capaldi2018probing, 
liu2022confinement} This mechanism was used to trap fluorescently stained DNA molecules into
the cavities formed by sealing off the nanopits. In some cases, single molecules were trapped,
and in other cases pairs of DNA molecules were trapped. Using a mixture of DNA molecules
stained with two different dyes, some cavities contained two DNA molecules that were visually 
distinguishable by colour, thus facilitating the observation and characterization of the 
organization and dynamics of the individual chains. The simplicity of the systems used in such 
{\it in vitro} experiments provide results that are much more amenable to direct quantitative 
comparison with computer simulations employing simple models than are the {\it in vivo} experiments 
studying bacterial chromosome segregation. 


The studies of Refs.~\onlinecite{capaldi2018probing} and \onlinecite{liu2022confinement} employed
cavities in which the DNA molecules are strongly confined in one dimension and much less so in the
lateral dimensions. In each case, the cavity height of 200~nm is appreciably smaller than the bulk
radius of gyration of the linear DNA molecules used (0.7~$\mu$m for $\lambda$ DNA), thus compressing
the chains in that dimension and enhancing repulsion in the lateral plane.
Ref.~\onlinecite{capaldi2018probing} employed cavities with a square cross section of side length
2~$\mu$m, while Ref.~\onlinecite{liu2022confinement} employed cavities with elliptical cross sections
of fixed cross-sectional area and variable eccentricity, $e$, with maximum diameter ranging from
2~$\mu$m ($e=0$) to 3~$\mu$m ($e=0.9$). Such values are comparable to the lateral extent of
singly-trapped $\lambda$ DNA molecules, thus ensuring that two
co-trapped molecules will interact strongly with each other.

In the case of square cavities,\cite{capaldi2018probing} the presence of a
second $\lambda$ DNA molecule had a pronounced effect on both the position probability
distribution and the chain dynamics. Entropic repulsion tended to push the molecules
apart and away from the cavity centre, the most probable location for single-chain trapping.
This effect was also observed, somewhat more weakly, when the $\lambda$ DNA chain was 
trapped with a much smaller plasmid. In addition, the $\lambda$-DNA dynamics were slowed
when a second chain was trapped, much more so in the case where the second chain is $\lambda$ 
DNA than for a plasmid. In the former case, the authors of Ref.~\onlinecite{capaldi2018probing}
infer a ``Brownian rotor'' collective motion of the two chains. They also
observed an asymmetry in the lateral centre-of-mass position distributions for
two-chain systems where the $\lambda$ DNA molecules are stained with different dyes. 
This likely arises because the two dyes (YOYO-1 and YOYO-3) unwind the double helix
by different amounts leading to different contour lengths.
In the case of elliptical cavities,\cite{liu2022confinement} elongating the cavity by
increasing its eccentricity increased the tendency for two co-trapped $\lambda$ DNA
molecules to segregate to the poles of the ellipse and decreased the rate with which
the two chains swap positions. Liu {\it et al.} also investigated the behaviour of a single
plasmid trapped with a T$_4$ DNA molecule in an elliptical cavity. They found 
that entropic repulsion with the T$_4$ DNA chain enhanced the probability of the plasmid lying
near the lateral walls, and increasingly so at the poles of the ellipse as the eccentricity 
was increased.

The nanocavity-confined DNA systems studied in Refs.~\onlinecite{capaldi2018probing} and 
\onlinecite{liu2022confinement} are simple enough to benefit from complementary studies
using computer simulations with simple molecular models. Recently,
we used Brownian dynamics and Monte Carlo simulation methods to study the organization and dynamics 
of two flexible Lennard-Jones chains trapped in cavities resembling those with square cross sections 
used in Ref.~\onlinecite{capaldi2018probing}.\cite{polson2021equilibrium} We calculated the
position probability distributions in the lateral plane and the centre-of-mass dynamics and
measured their variation with lateral box size and polymer length.  As in the experiments, the behaviour 
of the system was drastically altered by the presence of the second polymer, and was also highly 
sensitive to the cavity size. In smaller boxes the dynamical behaviour of the system was slowed 
due to the presence of the second polymer. The centre-of-mass dynamics were readily interpreted 
using the lateral position probability distributions. Generally, the results were qualitatively 
consistent with those of the experimental study of Capaldi~{\it et al.}\cite{capaldi2018probing}
However, the simulation results were not suitable for a direct quantitative comparison with
experiment. Because of the time-consuming nature of the dynamics calculations, the polymer chains 
were limited to lengths of $\lesssim 100$ monomers, which effectively results in an artificially large 
value of the polymer width relative to the persistence and contour lengths when the model is mapped 
onto $\lambda$ DNA. In addition, the presumed asymmetry in the contour lengths of the differentially
stained DNA molecules was not incorporated into the model.

In this study, we build on our previous work in Ref.~\onlinecite{polson2021equilibrium} and use 
Monte Carlo (MC) and Brownian dynamics (BD) simulations to study systems of two polymers entrapped 
in cavities of comparable shape and size as those used in the experiments. We address issues arising 
from the previous study by incorporating features that improve the semblance of the molecular model 
to the DNA systems studied in Ref.~\onlinecite{capaldi2018probing}. This includes using more realistic
values for the effective polymer width as well employing different polymer contour lengths for
each of the two polymers. Generally, 
we find that reducing the effective width also reduces the strength of interchain repulsion, thus
diminishing the tendency for polymer segregation. Confinement of polymers of different contour
length effects an asymmetry in the position probability distribution qualitatively similar
to that seen in the experiments. We also examine the effects of cavity elongation in the lateral
plane, employing rectangular cross sections as well as the elliptical cross sections used in
Ref.~\onlinecite{liu2022confinement}. As in the experiments, we find that for sufficiently
small cavities, increasing the elongation enhances segregation of the polymers along 
the long symmetry axis. In addition, the free energy barrier that governs chain swapping
grows, leading to a reduced swapping frequency, in accord with the experiments. 
A small ring polymer trapped in an elliptical cavity with large linear polymer tends to
locate near the poles of the ellipse as a result of enhanced entropic depletion of the
linear polymer in these regions, an effect also observed in experiments using a small
plasmid entrapped with a T$_4$ DNA molecule.\cite{liu2022confinement}

The remainder of this article is organized as follows. Section~\ref{sec:model} provides a 
description of the two polymer models employed in the MC and BD simulations. Section~\ref{sec:methods} 
gives an outline of the methodology used together with the relevant details of the 
simulations. Section~\ref{sec:results} presents the simulation results for the various systems.
Finally, Sec.~\ref{sec:conclusions} summarizes the main conclusions of this work.

\section{Model}
\label{sec:model}

In this study, we examine a system of two polymers confined to a box-like cavity.
The polymers are modeled as chains of spherical monomers. In most simulations, the 
chains are identical with $N$ monomers per chain and {are} of linear topology.  
In Sec.~\ref{subsec:AsymPA} we also consider polymers of unequal lengths $N_1$ and 
$N_2$, where we choose $N_2<N_1$, but which are otherwise identical. 
In Sec.~\ref{subsec:elliptical} we examine a system of two chains of unequal length
where one has a linear topology while the other is a ring topology.
Depending on whether we use MC or BD simulations, the chain is modeled as
either a semiflexible hard-sphere chain or a fully flexible chain of 
Lennard-Jones (LJ) monomers. The details of each are given below.

\subsection{Semiflexible hard-sphere chain}
\label{subsec:sfhs}
In most MC simulations we employ the semiflexible hard-sphere chain model. 
Here, the interactions between non-bonded monomers are given by
\begin{eqnarray}
u_{\rm nb}(r) =
\begin{cases}
 \infty & r < \sigma \\
 0,     & r \geq \sigma
\end{cases}
\label{eq:HSnb}
\end{eqnarray}
where $r$ is distance between centres of monomers. The bond length between sequentially adjacent 
monomers was {held fixed at a length} of $\sigma$. The width $w$ of the hard-sphere polymer 
is defined as the diameter of each monomer, i.e., $w=\sigma$.

The bending potential for each consecutive triplet of monomers centered on monomer $i$ is given by,
\begin{eqnarray}
    u_{\rm bend}=\kappa (1-\cos\theta_i),
    \label{eq:Rigid}
\end{eqnarray}
where $\cos\theta_{i}=\hat{u}_{i}\cdot\hat{u}_{i+1}$, and
where $\hat{u}_{i}$ is the unit vector pointing from monomer $i-1$ to monomer $i$. 
The bending constant $\kappa$ determines the overall stiffness of the polymer and 
is  related to the persistence length $P$ by\cite{micheletti2011polymers} 
$\exp(-\langle l_{\rm bond} \rangle/P) = \coth(\kappa/k_{\rm B}T) - k_{\rm B}T/\kappa$.  
For sufficiently large $\kappa/k_{\rm B}T$ this implies
$P/\sigma\approx \kappa/k_{\rm B}T$. In this limit, the Kuhn length, 
$\ell_{\rm k}=2P$, therefore satisfies $\ell_{\rm k}/\sigma\approx 
2\kappa/k_{\rm B}T$.

\subsection{Fully flexible Lennard-Jones chain}
\label{subsec:fflj}

In contrast to the hard-sphere chain model, this model employs continuous potentials 
and so is suitable for use in BD simulations and as well as MC simulations.
Here, the non-bonded interactions are given by the repulsive LJ potential,
\begin{eqnarray}
u_{\rm nb}(r) =
\begin{cases}
u_{\rm LJ}(r) - u_{\rm LJ}(r_{\rm c}),  & r \leq r_{\rm c} \\
0,   & r > r_{\rm c}
\end{cases}
\label{eq:LJ}
\end{eqnarray}
where  $u_\text{LJ}$ is the LJ potential,
\begin{equation}
u_\text{LJ}(r)= 4\epsilon((\sigma/r)^{12}-(\sigma/r)^6).
\label{eq:LJ2}
\end{equation}
Here, \textit{r} is the distance between the centres of the two interacting monomers and, 
\textit{$r_c$}=$2^{1/6}\sigma$, where $\sigma$ is the monomer diameter.
The interaction between bonded monomers is given by a combination of the LJ potential given in 
Eq. (\ref{eq:LJ}) and the Finitely Extensible Non-linear Elastic (FENE) potential given by
\begin{eqnarray}
u_\text{FENE}(r)= -\frac{1}{2}kr_0^2\ln(1-(r/r_0)^2),
\label{eq:FENE}
\end{eqnarray}
where \textit{$r_0$}=1.5$\sigma$, and $k\sigma^2/\epsilon=30$. The width $w$ of the
LJ polymer is defined as the diameter of each LJ monomer, i.e., $w=\sigma$.

\subsection{Confinement}
\label{subsec:confinement}

\begin{figure}[!ht]
\begin{center}
\vspace*{0.2in}
\includegraphics[angle=0,width=0.48\textwidth]{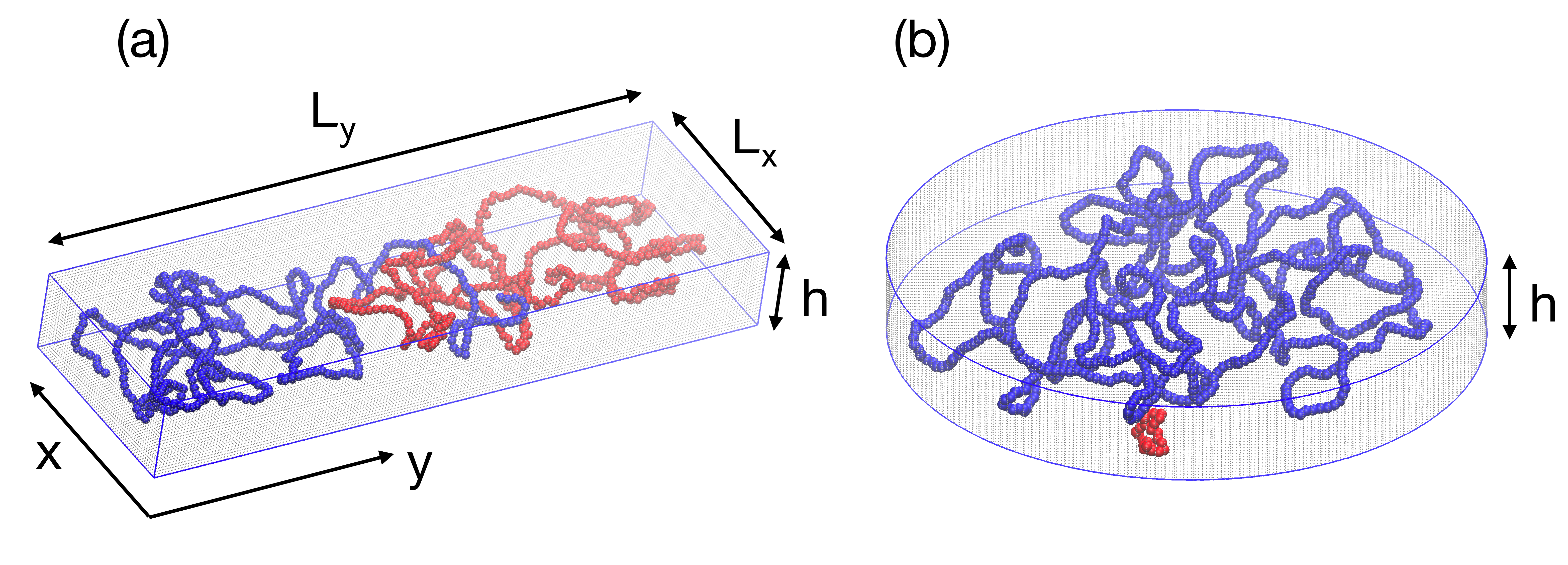}
\end{center}
\caption{{Illustration of the two-polymer model system. 
(a) Illustration of a system where the confining cavity is rectangular in shape. 
(b) Illustration of a system where the cavity has an elliptical cross section
(see Section~\ref{subsec:elliptical}.)
In both cases the distance $h$ between the top and bottom flat surfaces is constant.}}
\label{fig:illustration}
\end{figure}

In all systems, the polymers were confined to a cavity with cross-sectional area 
$A$ in the $x-y$ plane and constant height $h$ in the $z$-direction. In most calculations,
the cavity cross section was square or rectangular in shape, though elliptical
cross sections were also employed in a few simulations. In the former case,
the confining box-like cavity has lateral dimensions $L_x$ and $L_y$, where we 
choose  $L_x\leq L_y$. For rectangular cross sections ($L_x<L_y$), it is 
convenient to define the geometric average side length $\overline{L}\equiv 
\sqrt{L_xL_y}$. Monomer-wall interactions were modeled by means of a virtual monomer
placed at the nearest point on the wall from the actual monomer. The interaction
energy is calculated using. Eq.~(\ref{eq:HSnb}) in the case of the hard-sphere chain 
and by  Eqs.~(\ref{eq:LJ}) and (\ref{eq:LJ2}), in the case of the LJ chain. In each 
case, $r+\sigma$ is the distance between  the  centre of the monomer and the nearest 
point of interaction on the wall. Thus, the cross-sectional area, $A\equiv L_xL_y$, 
and height, $h$, define the space accessible to the centres of the monomers.
{The system is illustrated in Fig.~\ref{fig:illustration}.}

To facilitate a comparison between simulation results and experimental results, the
cavity dimensions used in the simulations are presented as ratios with respect to 
either the root-mean square (RMS) radius of gyration of the polymer in bulk, $R_{\rm g}$, 
or the RMS radius of gyration in the $x-y$ plane for a polymer confined to a slit 
of a given height, $R_{{\rm g},xy}$. The values of $R_{\rm g}$ and $R_{{\rm g},xy}$
for the DNA polymers employed in Refs.~\onlinecite{capaldi2018probing} and 
\onlinecite{liu2022confinement} have been measured experimentally under conditions
relevant to those cavity-confinement studies.

\section{Methods}
\label{sec:methods}
We use two different computer simulation methods to examine the 
equilibrium dynamical and non-dynamical behaviour of the system of two confined 
polymers. To investigate the dynamics of the centre-of-mass positions of the 
polymers we use BD simulations, and to examine the non-dynamical behaviour we use 
Metropolis MC simulations. The BD simulations can, in principle, provide the 
same  information as the MC simulations. However, the BD simulations are
significantly more time consuming, and thus we chose to use the MC simulations 
to obtain the  non-dynamical results. A brief description of the methods is given
below.

\subsection{Metropolis Monte Carlo simulations}
\label{subsec:mc}

The MC simulations employ standard techniques in which
a trial move is generated and accepted or rejected based on the Metropolis MC 
criterion. In simulations using the freely-jointed LJ chain 
model, we use three types of trial moves: (1) reptation moves, (2) crankshaft 
rotations, and (3) whole-polymer translation. In all simulations the type of 
trial  move was randomly chosen. Each MC simulation consisted of 
$N_{\rm tot}\equiv N_1 + N_2$ trial moves, where $N_1$ and $N_2$ are the
number of monomers in each polymer. Of these moves two were whole-polymer
translation,  $0.9(N_{\rm tot}-2)$ were 
reptation, and $0.1(N_{\rm tot}-2)$ were crankshaft rotations. In the simulations 
using the semiflexible hard-sphere chain model, only reptation and crankshaft
rotations were employed. {(The exception is for the system studied in
Section~\ref{subsec:elliptical}, where one of the polymers has a ring topology, 
in which case crankshaft and whole-polymer translation moves were used.)}
Typically, there were $0.9N_{\rm tot}$ reptation moves and 
$0.1N_{\rm tot}$ crankshaft rotations. Simulations using this model also 
employed a cell list to increase the efficiency of the program. In the 
case of crankshaft moves, a monomer was randomly selected and rotated about 
the axis  connecting its two neighbouring monomers by an angle that was drawn 
from a uniform  distribution on the interval [$-\phi_\text{max},\phi_\text{max}$]. 
The magnitude of  $\phi_\text{max}$ was chosen such that the acceptance rate 
was close to 50\% when  possible.  In the case of whole-polymer translation, 
the displacement of all monomers of a randomly selected polymer in each 
dimension was drawn from a uniform distribution on the interval 
[-$\Delta_\text{max},\Delta_\text{max}$]. The magnitude of $\Delta_\text{max}$ was 
chosen such that the acceptance rate was close to 50\%. 

In simulations where the polymers were modeled as freely jointed LJ chains, the 
polymers were each of length $N=60$ monomers.  For a calculations with given set of system 
parameter values, the results of 200 statistically independent simulations averaged. 
Individual simulations consisted of $10^6$ steps in the equilibration phase followed by a 
production phase of $10^8$ steps. These simulations typically ran for around 800 minutes.
Thus, each calculation required about 0.3 CPU years.

In simulations where the polymers were modeled as hard-sphere chains, the size of the 
polymers and box geometry varied. The calculation with the largest system size
($N_1=3300$ and $N_2=1900$) used 1000 statistically independent simulations, each consisting 
of an equilibration period of $10^5$ steps and a production period of $10^6$ steps. 
Individual simulations ran for about 2600 minutes, and so each result for a given
set of system parameter values required about 4.9 CPU years. 

For the MC simulation results presented in Sec.~\ref{sec:results}, distance
is measured in units of $\sigma (=w)$, where $\sigma$ is the hard-sphere
diameter in the case the semi-flexible hard-sphere chain model, and the LJ
particle diameter defined in Eq.~(\ref{eq:LJ2}) for the freely-jointed LJ chain
model. In addition, energy is measured in units of $k_{\rm B}T$.
Note that temperature chosen such that $\epsilon=k_{\rm B}T$ for all simulations
using the LJ chain model.

\subsection{Brownian dynamics simulations}
\label{subsec:bd}

The BD simulations used to study the polymer dynamics employ standard methods.
The coordinates of the {\it i}th particle are advanced through a time $\Delta t$ 
according to 
the algorithm:
\begin{eqnarray}
x_i(t+\Delta t) & = & x_i(t) + \frac{f_{i,x}}{\gamma_0} + \sqrt{2 k_{\rm B}T 
\Delta t/\gamma_0} \Delta w,
\label{eq:BDeq}
\end{eqnarray}
and likewise for $y_i$ and $z_i$.  Here, $f_{i,x}$ is the $x$-component of the 
conservative 
force on particle $i$, and $\gamma_0$ is the friction coefficient of each monomer. 
The conservative  force is calculated as $f_{i,\alpha}=-\nabla_{i,\alpha} 
U_{\rm tot}$, where $\nabla_{i,\alpha}$  is the $\alpha$-component of the gradient
with respect to the coordinates of the $i$th particle  of the total potential
energy of the system, $U_{\rm tot}$. In addition, $\Delta w$ is a random quantity
drawn from a Gaussian distribution of unit variance.  

BD simulations were only used for LJ chains in the rectangular cavities. The majority of 
results were obtained for $N=60$, though some calculations were conducted for chain lengths 
in the range of $N=40-70$. A result for a system with $N=60$ and selected values $\bar{L}$ 
and $R$ were obtained by averaging over 1000 statistically independent simulations, each 
consisting of a equilibration period of duration $\Delta t_{\rm eq}=10^4$ and a production 
period of $\Delta t_{\rm prod}=10^5$. An individual simulation required about 2100 minutes 
of CPU time, and thus the computational cost of each calculation was approximately 4 CPU 
years.

For the BD simulation results presented in Sec.~\ref{sec:results}, distance 
is measured in units of $\sigma (=w)$, energy is measured in units of 
$\epsilon (=k_{\rm B}T)$, and time is measured in units of 
$\gamma_0\sigma^2/\epsilon$.

\section{Results}
\label{sec:results}

\subsection{Long chains in cavities with symmetric cross sections}
\label{subsec:AsymCS}

In a recent study, we examined the organization and dynamics of two identical
polymers confined to a box with a square cross section using freely-jointed chains 
of ${\cal O}(10^2)$ monomers.\cite{polson2021equilibrium} Such short chains
were used because of the time consuming nature of the BD simulations.
The disadvantage of such a model is its inability to tune the ratios
of the relevant polymer length scales, i.e. the contour length $\ell_{\rm c}$, the 
Kuhn length, $\ell_{k}$, and the effective width, $w$, to values that all 
correspond to those of  $\lambda$-DNA used in the experiments. Most significantly,
the ratio  $\ell_{\rm k}/w=1$
of the model system is an order of magnitude lower than the experimental value.
As noted in Ref.~\onlinecite{polson2021equilibrium}, the artificially large 
effective width used in the simulations likely affects the observed physical 
behaviour of the polymers. The purpose of this section is to quantify this effect.
To do so, we examine semiflexible polymers with adjustable $\ell_{\rm k}$ and study 
the behaviour upon variation in the ratio $\ell_{\rm k}/w$ for various system
sizes. For realistic values of $\ell_{\rm k}/w$ and $\ell_{\rm c}/w$, the number
of monomer beads is ${\cal O}(10^3)$, which is impractically large to study
the dynamics. Consequently, we focus on the static property of the equilibrium
organization of the chains using MC simulations.

To reproduce the experimental conditions of  Ref.~\onlinecite{capaldi2018probing}, we
choose the effective width of the $\lambda$-DNA to be $w=$10~nm,\cite{tree2013dna} 
and a Kuhn length of $\ell_\text{k}=127.2$~nm.\cite{dobrynin2006effect} 
From Fig.~7 of Ref.~\onlinecite{kundukad2014effect} we estimate that a YOYO-1
staining ratio of 10:1 (bp:fluorophore) increases the contour length by about 15\%.
Thus, the contour length of $\lambda$ DNA increases from 16.5~$\mu$m\cite{wang1998scanning}
to $\approx 19~\mu$m.  To achieve this,
we used a bending coefficient of $\kappa$=6.36 and chain of $N$=1900 spherical
monomers. This yields a radius of gyration in the bulk of $R_{\rm g}=71.1\pm0.4$, 
and $R_{{\rm g},xy}=82.9 \pm 0.1$ in a slit of height $h=20w=200$~nm, consistent with the 
experiment.\cite{capaldi2018probing} With these parameters, we ran simulations 
for  three box sizes with scaled side lengths of 
$L/R_\text{g}$=1.38, 2.82 and, 4.20, where $L$ is the box width and where
$R_\text{g}$ is the bulk radius of gyration. These values 
correspond to boxes with lateral confinement sizes smaller than, equal to, and 
larger than the experimental values in Ref.~\onlinecite{capaldi2018probing}. 
We also consider polymers with twice the effective thickness, i.e.,
$\ell_{\rm k}/w$ = 6.36, and the same contour length by reducing
both $\kappa$ and $N$ by a factor of two. Finally, we compare with a freely-jointed
chain with $\ell_{\rm k}/w$=1 with the same $\ell_{\rm c}/w$ ratio. In each
case, the number of beads $N$ is adjusted to maintain a fixed ratio $\ell_{\rm c}/w$,
and the box height eight is fixed at $h/R_\text{g}=0.282$, as it was in the 
experiments. Note that $w$ is of course tunable in experiments by variation of
the ionic strength.

\begin{figure}[!ht]
\begin{center}
\vspace*{0.2in}
\includegraphics[angle=0,width=0.48\textwidth]{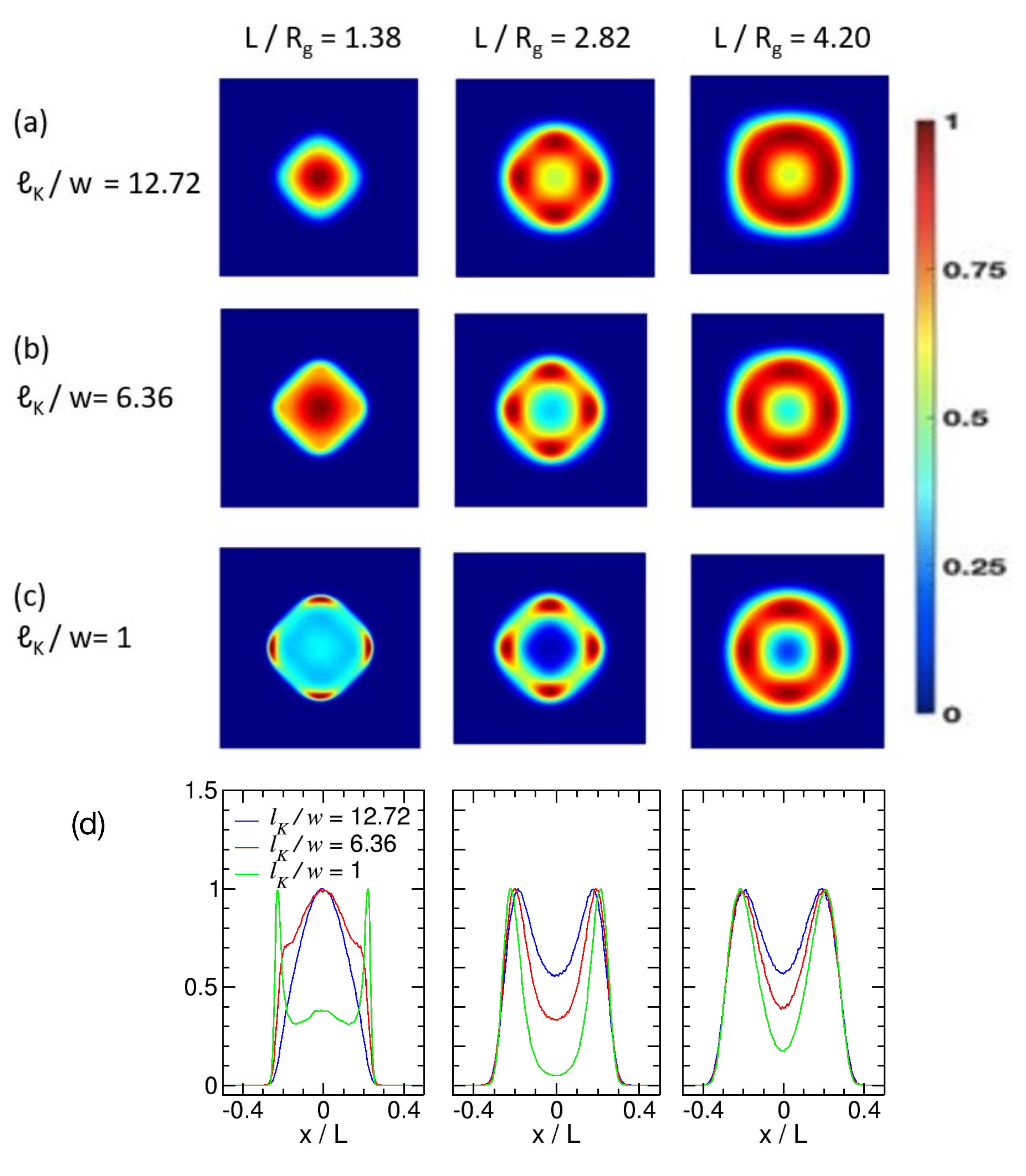}
\end{center}
\caption{(a)--(c) Polymer COM probability distributions, ${\cal P}(x,y)$, for a system two  
hard-sphere chains confined to cavities with square cross sections of 
side length $L$.  The cavities are scaled such that $h/R_\text{g}=0.282$ and 
the polymers each have a fixed number of Kuhn segments given by
$\ell_\text{c}/\ell_\text{k}$=149. Each row corresponds to a different value of 
$\ell_{\rm k}/w$, and each column corresponds to a different scaled box width,
 $L/R_{\rm g}$.  Distributions in rows (a) and (b) were calculated using semiflexible 
polymer chains, while those in row (c) used freely-jointed chains. (d) Cross sections 
of the distributions through $y=0$.}
\label{fig:width}
\end{figure}

Figure~\ref{fig:width} shows centre-of-mass (COM) probability distributions
for either polymer in the $x-y$ plane, ${\cal P}(x,y)$. The distributions in
row (a) show results for systems with a value of $\ell_{\rm k}/w$ that is
approximately equal to the experimental value. Those in rows (b) and (c) have smaller
values of $\ell_{\rm k}/w$. Since $\ell_{\rm c}/\ell_{\rm k}$ is fixed, these
distributions in effect correspond to polymers of larger polymer width $w$.
The distributions in row (c) were calculated using freely-jointed chains, as in
our previous study.\cite{polson2021equilibrium} The distributions in each column
correspond to three different values of $L/R_{\rm g}$.
The graphs in row (d) are cross sections of the distributions for
$y=0$.  The distributions in the middle column were calculated for a box width of
$L/R_{\rm g}$=2.82, which is approximately the value in the experiments.
Thus, the columns on the left and right correspond to box widths smaller and
larger than the experimental values, respectively. Note that the distribution
calculated using conditions that best approximate those of the experiment
of Capaldi {\it et al.}\cite{capaldi2018probing} is that shown the middle panel
of row (a). 
 
Consider first the middle column, where $L/R_{\rm g}$=2.82. For all $\ell_{\rm k}/w$,
the distributions show roughly the same qualitative features: a depletion zone
in the middle of the box, and a ring-like structure containing four local
high-probability spots located at the midpoints of the box edges. As noted in 
Ref.~\onlinecite{polson2021equilibrium}, entropic repulsion between the confined
polymers tends to push them to opposite sides of the box. {Thus, when} 
the COM of one polymer is located at position $(x,y)$, the other tends to be at $(-x,-y)$, 
with the square box shape tending to favour positions at the box-edge midpoints. 
The depletion zone at the box centre results from the high entropic cost of 
polymer overlap for this box size. As $w$ increases, the packing fraction
inside the cavity increases as well, increasing entropic repulsion between
the chains and enhancing segregation.  This appears as enhanced depletion
in the box centre and an increased intensity of the probability ring. This is more 
clearly evident in the probability cross section graph in the middle panel of row (d).
For the widest chains with $\ell_{\rm k}/w=1$, the polymer centre tends to further 
localize to four ``hot spots''. Thus, high packing fraction promotes greater
positional ordering in addition to segregation. 

The effects of varying the cavity width are somewhat complicated. In the case of
semiflexible chains ($\ell_{\rm k}/w=12.72$ and $\ell_{\rm k}/w=6.36$) there is 
a small enhancement in depletion at the box centre as the cavity width decreases from
$L/R_{\rm g}=4.20$ to $L/R_{\rm g}$=2.82. This implies a slight increase in the
degree of chain segregation. However, in decreasing the cavity width further to 
$L/R_{\rm g}=1.38$, lateral confinement overrides entropic repulsion, yielding
distributions with polymer centers strongly localized in the middle
of the cavity. Increasing chain thickness $w$ enhances chain repulsion, 
leading to a slight broadening of the central peak for $\ell_{\rm k}/w=6.36$ relative
to that for $\ell_{\rm k}/w=12.72$. For the 
widest chains ($\ell_{\rm k}/w=1$) there is a similar enhancement in depletion at
the box centre as $L$ decreases from $L/R_{\rm g}=4.20$ to $L/R_{\rm g}$=2.82,
though the change is more pronounced than for semiflexible chains. Likewise, for 
the smallest cavities with $L/R_{\rm g}=1.38$ confinement overcomes the entropic 
repulsion, leading to an increase in the probability at the cavity center. In this 
extreme case of thick chains confined to a very small cavity,
the combination of high packing fraction and lateral confinement 
yields a qualitatively distinct distribution with a broad central plateau with four 
high-probability peaks at positions comparable to those for the case of larger $L$. 

To summarize, varying the polymer width and lateral confinement both strongly affect
the equilibrium organization of the polymer chains. Generally, increasing chain
width enhances entropic repulsion between chains and promotes segregation, while
the effects of varying the degree of confinement are somewhat more complex.
A partial depletion of DNA COM probability in the cavity centre was also observed 
in the experiments, consistent with the result for the system that best approximates
the experimental conditions. Here, the observed ring-like structure is consistent with 
the interpretation of the dynamical behaviour of the $\lambda$ DNA molecules as collective
Brownian rotation about the box centre. The sharp peaks observed for $\ell_{\rm k}/w$=1 
and small boxes are spurious features produced by the unphysically high volume fraction 
for such polymer widths and are unlikely to be relevant to the systems studied in 
Ref.~\onlinecite{capaldi2018probing}. For the experimentally relevant box size
($L/R_{\rm g}=2.82$), using artificially wide polymer chains in simulations gives
rise to features in ${\cal P}(x,y)$ that will likely impact the dynamics in simulations.
Still, the gross features are sufficiently similar that the qualitative trends are
expected to be similar except for the case of very small boxes. 
 
\subsection{Asymmetry in polymer length}
\label{subsec:AsymPA}

In this section we examine the physical behaviour of two confined polymers of 
different contour lengths. One motivation for considering such a
system is provided by the experiments of Ref.~\onlinecite{capaldi2018probing}.
In that study two confined $\lambda$-DNA strands were stained with 
different dyes (YOYO-1 and YOYO-3) in order to distinguish between them. Unlike the 
COM probability distributions presented in Sec.~\ref{subsec:AsymCS}, there was
a clear asymmetry in the experimental distributions. In particular, the DNA strand 
stained with YOYO-3 tended to lie closer to the centre of the box, while
that stained with YOYO-1 was somewhat more displaced from the centre.
This asymmetry most likely arises because the two dyes unwind the dsDNA by
differing amounts, leading to slightly different contour lengths.  In addition
to that system, Capaldi {\it et al.} also examined a $\lambda$-DNA 
strand confined with a DNA plasmid, which has a much smaller contour length than 
$\lambda$-DNA. We anticipate that future experiments will extend this work and examine 
a larger number of contour length ratios. The simulation results presented in this 
section should be helpful for interpreting the results of such experiments. As before, 
we consider only confinement in a cavity with a square cross section, as in 
Ref.~\onlinecite{capaldi2018probing}. 

\begin{figure*}[!ht]
\begin{center}
\vspace*{0.2in}
\includegraphics[angle=0,width=0.8\textwidth]{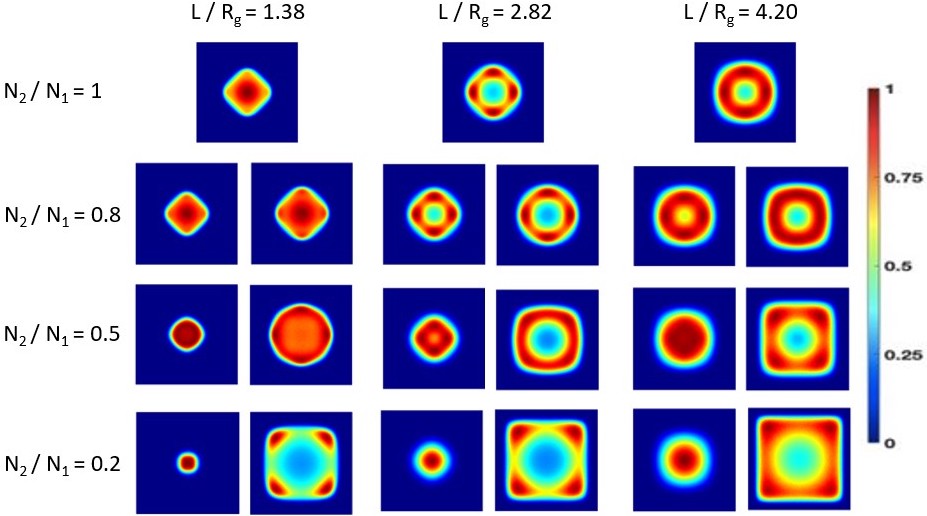}
\end{center}
\caption{COM probability distributions, ${\cal P}(x,y)$, for a system of two
semi-flexible hard-sphere chains of different lengths for various box widths and
polymer-length aspect ratios. In these results $N_1=950$ and $N_2$ is variable. 
In addition, $h$=10.71 and $\ell_\text{k}=6.36$. The left-most graph in
the pair of graphs in each column is the distribution for the longer  polymer,
while the right graph is that for the shorter polymer. }
\label{fig:prob2d_N1_95}
\end{figure*}

Figure~\ref{fig:prob2d_N1_95} shows COM probability distributions ${\cal P}(x,y)$ 
for a system two semiflexible  hard-sphere chains, one with length $N_1=950$ monomers
and the second of variable length with $N_2\leq N_1$. Each chain has a Kuhn length of
$\ell_\text{k}/w=6.36$. Results are shown for confinement to a box of height
$h=10.71$ for three different values of the scaled cavity width, $L/R_{{\rm g}}$. 
For each case, the distributions for the two polymers become more dissimilar 
as the ratio $N_2/N_1$ decreases 
from unity, as the longer polymer tends to be more concentrated at the centre while the 
shorter one is pushed further from the centre. Notably, in cases of a large
polymer length asymmetry, the  shorter polymer tends to be pushed to the 
corners of the box. By contrast, for smaller length asymmetry, the centre-of-mass 
probability tend to be somewhat more focused at the midpoints of the sides of the box.

A theoretical calculation designed to reproduce all of the trends in 
Fig.~\ref{fig:prob2d_N1_95} would be challenging and is beyond the scope of the
present work. Here, we pursue instead the more modest goal of elucidating the behaviour
in the case of large polymer length asymmetry (i.e. small $N_2/N_1$) and high packing
fraction (i.e. low $L/R_{\rm g}$). We follow the approach taken in the recent study of
Liu {\it et al.} in modeling the behaviour of a small plasmid in the presence a large DNA 
molecule in a nanoscale cavity.\cite{liu2022confinement} When the shorter polymer is 
sufficiently small, it is not expected to significantly perturb the monomer density
of the larger polymer, $\rho_{\rm mon}(x,y)$. In this limit the interpolymer
interaction free energy is approximately $F_{\rm int}=a \rho_{\rm mon}(x,y)$, where
$(x,y)$ is the COM position of the small polymer and $a$ is a constant. [Note that
the monomer density is defined to satisfy the normalization $\int_{\cal A} \rho_{\rm mon}(x,y) 
dx dy = N_1$, where the integral is over the area of the cross section.]  
The short polymer also interacts with the lateral walls with a 
free energy $F_{\rm wall}$ that depends on the distance to each wall. We used a simulation 
of a single confined polymer to directly calculate $\rho_{\rm mon}(x,y)$. In addition,
we carried out a set of simulations to measure the variation of $F_{\rm wall}$ with COM distance 
from a single lateral wall using a multiple-histogram technique.\cite{frenkel2002understanding}
[See Appendix~B of Ref.~\onlinecite{polson2019free} for details of a comparable
calculation.]

\begin{figure}[!ht]
\begin{center}
\vspace*{0.2in}
\includegraphics[angle=0,width=0.45\textwidth]{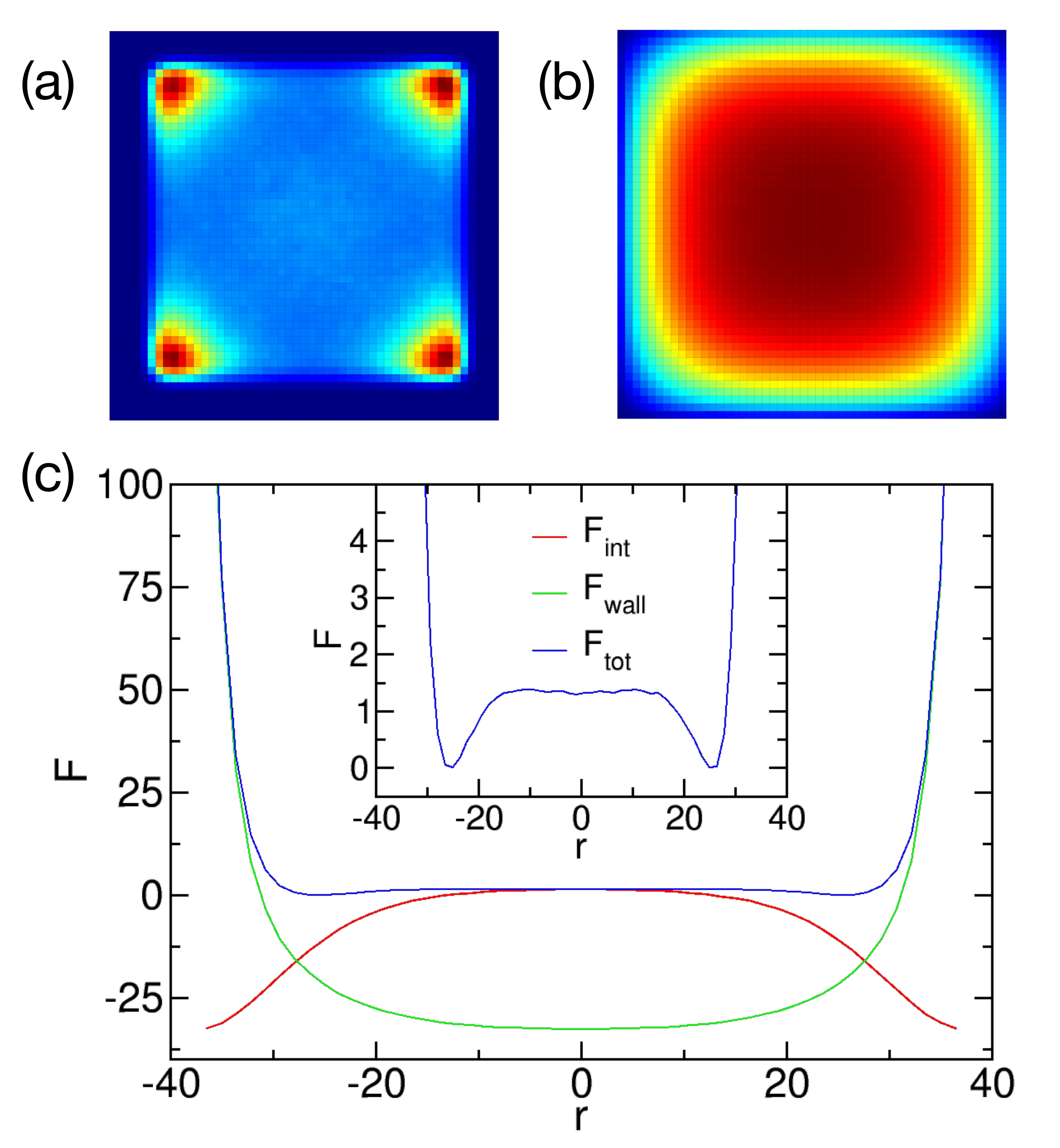}
\end{center}
\caption{(a) Theoretical calculation for the COM distribution ${\cal P}(x,y)$ of the short 
polymer for a system with $L/R_{\rm g}=1.38$, $N_2/N_1=0.2$ and $N_1=950$. (b) Monomer 
density, $\rho_{\rm mon}(x,y)$ for a single polymer of length $N=950$ confined to the same 
box as in (a).  (c) Variation of the free energy of the two-polymer system with distance 
along the diagonal of the box in the $x-y$ plane. $F_{\rm wall}$ is the free energy arising 
from interactions of {the short} polymer with the walls,  $F_1$ is 
the interpolymer interaction free energy, and $F_{\rm tot}$ is the sum of the two. 
{For convenience, $F_{\rm tot}$ has been vertically shifted so that
$F_{\rm tot}=0$ at the minima.} The inset shows a close-up of $F_{\rm tot}$.
}
\label{fig:prob_theory}
\end{figure}

Figure~\ref{fig:prob_theory}(a) shows
the resulting probability distribution. The results are qualitatively consistent with
the corresponding simulation result for the short polymer shown in Fig.~\ref{fig:prob2d_N1_95}
for $N_2/N_1=0.2$ and $L/R_{\rm g}=1.38$.  The four probability peaks in the corners of
the box are present. Figure~\ref{fig:prob_theory}(b) shows the monomer density
for the larger polymer $\rho_{\rm mon}(x,y)$. Note the depletion zones near the edges of
the box are enhanced near the corners, and it is this feature that tends to drive
the smaller polymer into these locations. The interpolymer repulsion is counterbalanced
by entropic repulsion of the small polymer with the walls. Figure~\ref{fig:prob_theory}(c)
shows how the contributions of $F_{\rm int}$ and $F_{\rm wall}$, here measured as a function
of displacement along the diagonal of the box, lead to the appearance of local free energy
minima corresponding probability peaks in Fig.~\ref{fig:prob_theory}(a). 
As a result of the approximations employed the results are only qualitatively consistent
with the simulation results. These calculations used a coupling value of $a/w^2=0.0236$, 
where $w$ is the polymer width and chosen unit of length (i.e., $w=1$). This value was
chosen to yield the same ratio for ${\cal P}(x,y)$ at the peaks relative to that at the
box centre ($x=y=0$) as in the simulation result in Fig.~\ref{fig:prob2d_N1_95} for
$N_2/N_1=0.2$ and $L/R_{\rm g}=1.38$.

\begin{figure}[!ht]
\begin{center}
\vspace*{0.2in}
\includegraphics[angle=0,width=0.45\textwidth]{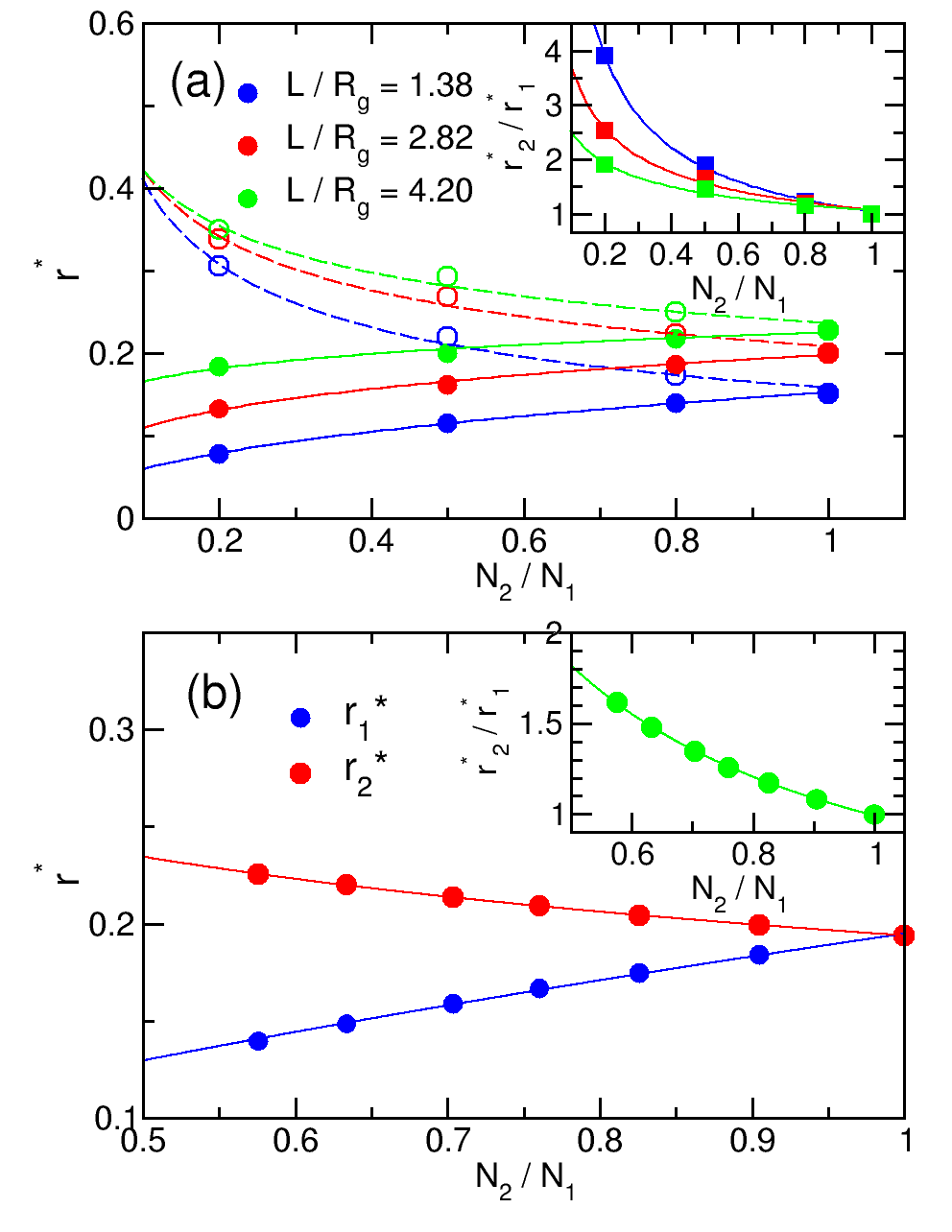}
\end{center}
\caption{(a) Scaled RMS radial distance of each {polymer} from the centre of the box, 
$r_1^*$ (solid symbols) and $r_2^*$ (open symbols), vs the polymer length ratio, $N_2/N_1$. 
Results are shown for various box widths. The two chains each have a scaled Kuhn length of 
$\ell_\text{k}/w=6.36$, and the long polymer has a length of $N_1=950$. In all cases, the 
box height is $h$=10.71.  The inset shows the ratio $r_1^*/r_2^*$ vs $N_2/N_1$. The solid 
and dashed lines are guides for the eye. (b) As in panel~(a), except for fixed
$N_2=1900$ and variable $N_1$. In addition, the Kuhn length is $\ell_\text{k}/w=12.72$, and 
the cavity dimensions are $h$=20 and $L$=200. }
\label{fig:radial}     
\end{figure}

The distributions in Fig.~\ref{fig:prob2d_N1_95} can be characterized using
the scaled RMS centre-of-mass displacement from the centre of the box, defined as
\begin{equation}
r_i^*\equiv\frac{\sqrt{\langle X_i^2+Y_i^2\rangle}}{L},
\label{eq:ristar}
\end{equation}
where $X_i$ and $Y_i$ are the COM coordinates of polymer $i$. The results are plotted 
in Fig.~\ref{fig:radial}(a) for a system of two polymers confined to a box with a height 
of $h=$10.71 for various polymer length ratios and box sizes. In each case, $N_1=950$
and $\ell_\text{k}=6.36$. For each box size the shorter polymer is increasingly displaced 
from the centre of the cavity as  $N_2/N_1$ decreases while the longer polymer is 
increasingly pushed toward the centre. This is consistent with the trends in
Fig.~\ref{fig:prob2d_N1_95}. The inset in Fig.~\ref{fig:radial}(a) shows the ratio 
$r_2^*/r_1^*$ vs the polymer length ratio. Note that $r_2^*/r_1^*$ is a measure of 
asymmetry in the probability distributions, with increasing asymmetry as $r_2^*/r_1^*$ 
increases from unity.  Thus, the results show that the distribution asymmetry increases 
with increasing lateral confinement. 

In the experiments of Refs.~\onlinecite{capaldi2018probing} and \onlinecite{liu2022confinement}
Capaldi {\it et al.} examined the case of a small DNA plasmid confined with a much
larger $\lambda$ DNA and observed that the plasmid was pushed out from the centre 
of the box while the larger one was more concentrated at its centre. 
This is qualitatively consistent with our results for small $N_2/N_1$. Unfortunately, their 
sampling was insufficient to determine if there was any preference for the plasmid to
locate to regions near the corners of the box. They further noted an asymmetry in the distributions 
of two differently stained $\lambda$-DNA molecules, suggesting that the stains change the 
contour length of the molecules by different amounts. They speculated that the longer polymer 
is pushed to the edges as it would be more easily deformed. By contrast, our simulation
data demonstrates it is instead the shorter polymer that tends to locate closer to the lateral
walls. 

\begin{figure*}[!ht]
\begin{center}
\vspace*{0.2in}
\includegraphics[angle=0,width=0.85\textwidth]{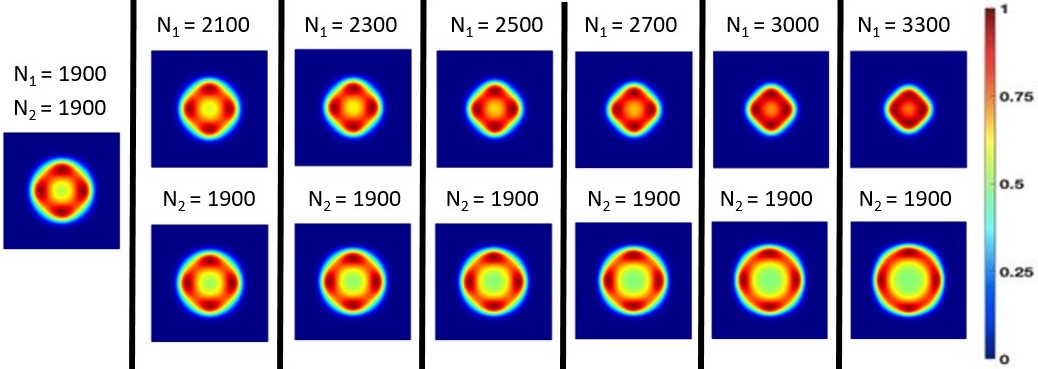}
\end{center}
\caption{Probability distributions, ${\cal P}(x,y)$, for a system two 
semiflexible hard-sphere chains with $\ell_\text{k}=12.72$, {
$N_\text{2}$=1900 and various $N_\text{1}$}. 
The polymers are confined to a box with a square cross section and dimensions of $L=200$ and $h=20$.}
\label{YOYO-3est}
\end{figure*}

Staining DNA with dyes such as YOYO-1 has the effect of partially unwinding the double 
helix and increasing the contour length. While such effects have been well characterized
for YOYO-1,\cite{kundukad2014effect} those of YOYO-3 have been less studied. Quantifying the 
asymmetry between the two distributions presented in Ref.~\onlinecite{capaldi2018probing} 
could in principle provide a means to estimate the contour length change caused by
the latter dye. To explore this option we examined a two-polymer system in which one 
polymer has properties of $\lambda$ DNA stained with YOYO-1 under conditions corresponding 
roughly to those of the experiments of Ref.~\onlinecite{capaldi2018probing}.  As in 
Sec.~\ref{subsec:AsymCS}, we estimate that a YOYO-1 staining ratio of 10:1 (bp:fluorophore) 
increases the $\lambda$ DNA contour length by about 15\% length from 16.5~$\mu$m to 
$\approx 19~\mu$m. Choosing an effective DNA width of $w=10~\mu$m, this corresponds to
$\ell_{\rm c}/w=1900$. Consequently, we use a semiflexible polymer of length $N_2=1900$ 
monomers. In addition, we use a Kuhn length of $\ell_{\rm k}/w=12.72$. To model $\lambda$ DNA
stained with YOYO-3, we assume this dye does not appreciably affect the Kuhn length or 
the effective width of the DNA, as is the case with YOYO-1.\cite{kundukad2014effect} 
Since our simulation results 
suggest that DNA strands stained with YOYO-3 are longer than those stained with YOYO-1,
we carry out a set of simulations for polymer lengths $N_1 \geq N_2$. Finally, we use
box dimensions of $L/w=200$ and $h/w=20$, which correspond to cavity the dimensions of 
$L=2$~$\mu$m and $h=200$~nm employed in Ref.~\onlinecite{capaldi2018probing}.

Figure~\ref{YOYO-3est} shows COM position probability distributions, ${\cal P}(x,y)$,
for various values of $N_1$. As expected, the distribution for the longer polymer is more 
concentrated at the centre than is the case for the shorter polymer for all values of
$N_1$. In addition, the distribution asymmetry grows with increasing polymer length
asymmetry. Figure~\ref{fig:radial}(b) shows the variation of $r_1^*$ and $r_2^*$ 
(defined in Eq.~(\ref{eq:ristar})) as well as the ratio, $r_2^*/r_1^*$, with respect
to $N_1$. The results in Ref.~\onlinecite{capaldi2018probing} were not presented in a
manner to facilitate a straightforward comparison with the simulation results, and
there may be insufficient spatial resolution to compare the distributions anyway.
However, in principle it should be simple to calculate quantities like $r_1^*$ and $r_2^*$ 
from the experimental data. We anticipate that such analyses in future experimental
studies in conjunction with these and future simulation results will provide a means
to quantify the effect of YOYO-3 and other dyes on the physical properties of DNA.

\subsection{Anisometry in box geometry}
\label{subsec:AsymBG}

Thus far, we have considered the behaviour of two polymers confined to a box-like cavity
with a square cross section, in accord with the experiments of Ref.~\onlinecite{capaldi2018probing}.
In a more recent study, Liu {\it et al.} \cite{liu2022confinement} examined the confinement 
in an elongated cavity with an elliptical shape in the lateral plane. As in 
Ref.~\onlinecite{capaldi2018probing} they consider the case of two DNA strands of comparable
length as well that of a long DNA strand confined with a much smaller plasmid. As in 
Ref.~\onlinecite{capaldi2018probing}, they examine the equilibrium organization 
and the dynamics of the confined molecules and show that both sets of properties are 
strongly affected by cavity elongation. Although other simulation studies have
examined the effects of cavity elongation for similar two-polymer 
systems,\cite{jun2010entropy,jung2012intrachain,polson2018segregation}
to our knowledge, none have employed the type of confining geometry used in 
Ref.~\onlinecite{liu2022confinement}, which is characterized by very strong confinement
in one dimension between flat surfaces. The purpose of this section (as well as the
following section) is to carry out simulations using a confining geometry that is
relevant to such experiments. We first consider the simple case of two identical
polymers confined to elongated cavities with a rectangular cross section and
examine both the configurational statistics and the equilibrium dynamics.
In the subsequent section we turn to cavities with elliptical cross sections
and consider polymers of equal and unequal contour lengths.

As noted in our previous study\cite{polson2021equilibrium} the simulations employed
to study polymer dynamics are much less efficient than the MC simulations used
to examine the configurational statistics. Consequently, systems using much shorter
polymers are required in the former case for computational feasibility. As a key goal
of the present work is to relate the equilibrium dynamics to the underlying free energy 
landscape of the system, the probability distributions for COM positions for such short-chain 
systems are needed. However, as noted for the case of cavities with square sections
in Sec.~\ref{subsec:AsymCS} any attempt to model the experimental system with short 
freely-jointed chains leads to distributions that may differ significantly from those
expected for the more realistic molecular models. Consequently, it is of value to 
first quantify this effect.

\begin{figure}[!ht]
\begin{center}
\vspace*{0.2in}
\includegraphics[angle=0,width=0.45\textwidth]{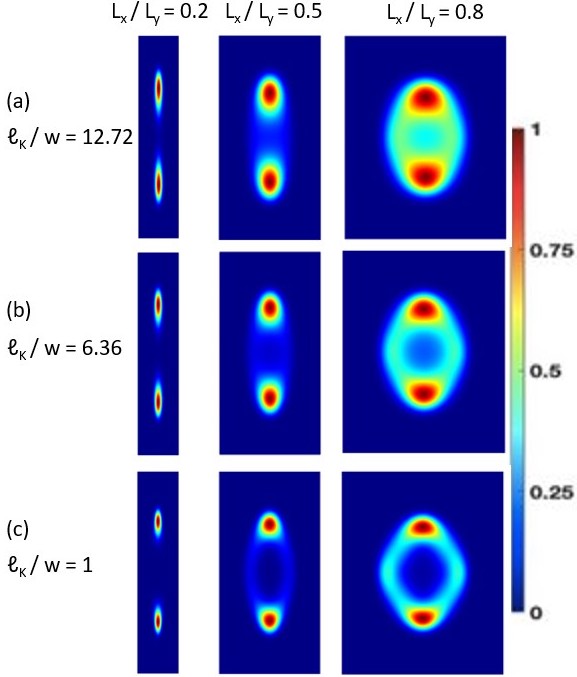}
\end{center}
\caption{Probability distributions for a system two semiflexible hard-sphere chains with 
various widths confined to cavities with rectangular cross sections.  The cavity dimensions 
satisfy $\overline{L}/R_{\text{g},xy}=2.82$, $\bar{L}/h=10$. For all $\ell_\text{k}/w$ values, the contour 
length is fixed to $\ell_{\rm c}/\ell_\text{k}=149$.}
\label{fig:widthrectangle}
\end{figure}

Figure~\ref{fig:widthrectangle} shows the effects of varying the polymer width on the
COM distributions of two identical polymers confined to a box of lateral dimensions
$L_x$ and $L_y (\geq L_x)$. In each case, the cross-sectional area is fixed such
that $\overline{L}/R_{\text{g},xy}=2.82$, where $\overline{L}\equiv \sqrt{L_xL_y}$
is the geometric mean lateral box length.  In addition, we fix the ratios 
$h/R_\text{g}=0.282$, and $\ell_{\rm c}/\ell_\text{k}=149$, where $\ell_{\rm c}$ and
$\ell_\text{k}$ is the contour length and Kuhn length, respectively.
Results for three different values of $\ell_\text{k}/w$ are shown, each for three
different cavity dimension asymmetries. 
Figures~\ref{fig:widthrectangle}(a) and (b) show results for semiflexible chains of
two different Kuhn lengths for model systems with properties comparable to those
studied in the experiments. In each case, cavity elongation dramatically
increases the tendency for the polymers to localize at two positions along the 
vertical dimension.  The main qualitative effect of decreasing $\ell_{\rm k}/w$
is the enhancement of the depletion in the centre of the box, most notably for 
$L_x/L_y=0.8$. For the case of $\ell_{\rm k}/w=1$ in Fig.~\ref{fig:widthrectangle}(c), 
this depletion is even more marked. In addition, the size of the two localized regions
decreases. Similar results were noted in Sec.~\ref{subsec:AsymCS} for square cross
sections. The results indicate that using an artificially large polymer width increases
the tendency for the polymers to segregate. 

\begin{figure*}[!ht]
\begin{center}
\vspace*{0.2in}
\includegraphics[angle=0,width=0.825\textwidth]{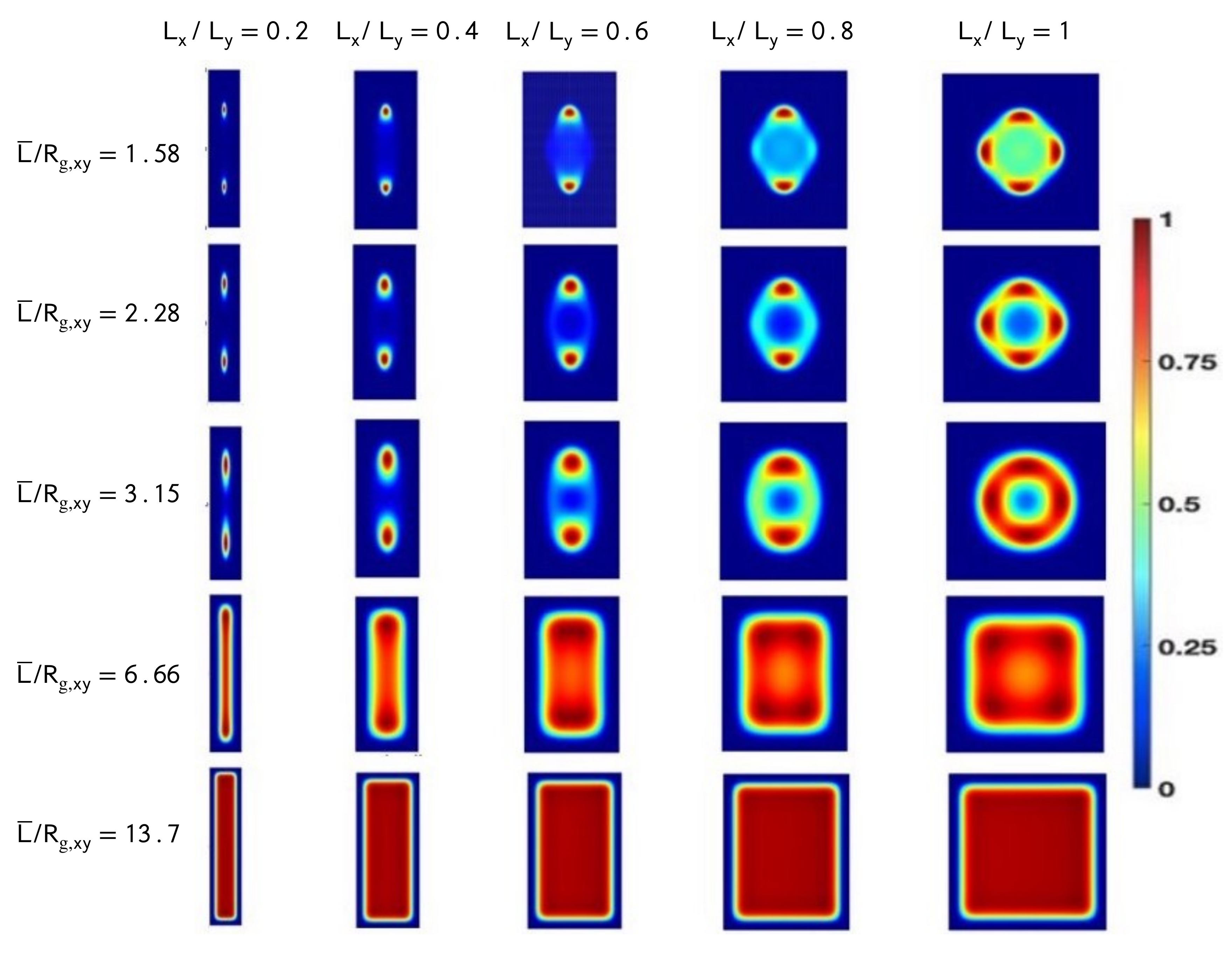}
\end{center}
\caption{COM probability distributions for a system two fully flexible LJ chains of 
length $N$=60 confined to a box of height $h$=4 and a rectangular cross sections. Results are 
shown for various values of the scaled cavity width, $\overline{L}/R_{{\rm g},xy}$, 
and box aspect ratio, $L_x/L_y$. }
\label{fig:prob2d_N60}
\end{figure*}

Figure~\ref{fig:prob2d_N60}  shows centre-of-mass position probability distributions 
in the lateral plane for two freely-jointed LJ chains, each of length $N$=60, confined to a
cavity a box height of $h=4$. Note that the LJ chain model is also amenable for use
in dynamics simulations, which are indeed used later in this section to characterize
the equilibrium dynamics for chains of this length. Results are shown for several values 
of the geometric mean box width, $\overline{L}$, as well as for various values of $L_x/L_y$. 
The trends for varying $L_x/L_y$ are qualitatively similar to those of Fig.~\ref{fig:widthrectangle}, 
which were calculated for hard-sphere chains. As the box becomes more elongated, i.e., as $L_x/L_y$ 
decreases, the polymers tend to segregate along the long axis of the box for all box sizes. 
The exception is for $\overline{L}/R_{{\rm g},xy}=13.66$, where the box is so large the 
polymers only rarely interact with each other.

To better characterize the effects of cavity anisometry on these distributions, we employ 
two different measures of the centre-of-mass displacement from the box centre, located at 
{$(x,y)=(0,0)$}, each in both the $x$ and $y$ directions. The first is defined
\begin{eqnarray}
\xi_x \equiv \sqrt{\langle X^2\rangle/\langle X_{\rm sq}^2\rangle},
\label{eq:xidef}
\end{eqnarray}
where $X$ is the centre-of-mass position of either polymer in the $x$-direction.
In addition, $X_{\rm sq}$ {is} the COM position for a box {with}
 a square cross section, where $L_x/L_y=1$. The quantity $\xi_y$ is likewise defined for the 
$y$ direction.  The second measure is defined
\begin{equation}
X^* \equiv {\sqrt{\langle X^2\rangle}}/{\overline{L}}.
\label{eq:Xstardef}
\end{equation}
The quantity $Y^*$ is similarly defined for the $y$-direction. 

\begin{figure}[!ht]
\begin{center}
\vspace*{0.2in}
\includegraphics[angle=0,width=0.45\textwidth]{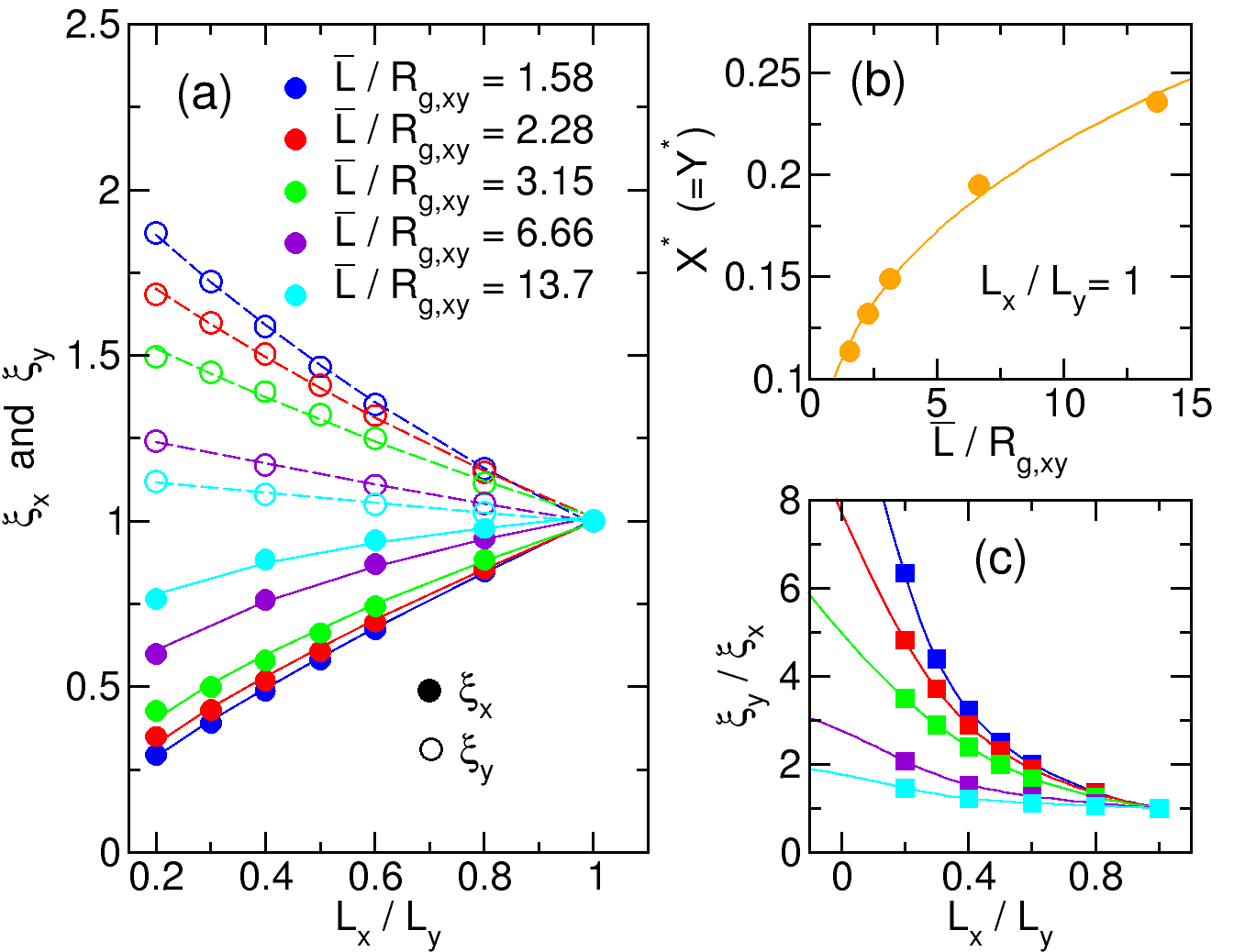}
\end{center}
\caption{(a) Variation of $\xi_x$ and $\xi_y$ (defined in Eq.~(\ref{eq:xidef})) with box 
anisometry, $L_x/L_y$. Results are shown for various box sizes for a system of two fully-flexible 
LJ chains with $N$=60 confined to a box of height $h=4$. (b) Variation of $X^*$ ($=Y^*$) 
(defined in Eq.~(\ref{eq:Xstardef})) with {scaled} box size for square boxes.  
(c) Ratio $\xi_x/\xi_y$ {vs} box anisometry using the results shown in panel~(a).} 
\label{fig:RMS-rectanglebox}
\end{figure}

Figure~\ref{fig:RMS-rectanglebox}(a) shows $\xi_x$ and $\xi_y$ vs $L_x/L_y$ for various
box sizes. As the box elongates (i.e., $L_x/L_y$ decreases from unity), $\xi_x$ decreases,
indicating that the distributions narrow along the $x$ direction, i.e. along the short
lateral dimension. This effect is enhanced by reducing the box size, i.e., the 
{rate} of 
decrease in $\xi_x$ from unity is greater as $\overline{L}$ decreases because the
depletion layer near the walls takes up a greater fraction of the box width along $x$.
By contrast, $\xi_y$ increases, indicating that the polymers are each 
displaced farther away from the box centre in the $y$-direction, consistent 
with the increased tendency for polymer segregation along this axis and the increasing
distance between probability peaks in Fig.~\ref{fig:prob2d_N60}.
This effect is enhanced by reducing the box size, i.e., the increase of 
$\xi_y$ from unity is greater as $\overline{L}$ decreases. For such smaller boxes, 
the polymers are forced into contact with each other, and the only means to prevent
overlap is for each to avoid the box centre.
 
Figure~\ref{fig:RMS-rectanglebox}(b) shows the variation of $X^* (=Y^*)$ with box size 
for the case of square cavities with $L_x/L_y=1$. The decrease of $X^*$ with 
$\overline{L}/R_{{\rm g},xy}$ is due to the increasing fraction of the box interior
occupied by the entropic depletion zones near the walls and corners as the cavity
size is reduced. This effect inhibits the polymers from moving too far from the centre.  
Finally, Fig. \ref{fig:RMS-rectanglebox}(c) shows the variation of the ratio  
$Y^*/X^*$ ($=\langle Y^2\rangle/\langle X^2\rangle$) with box anisometry, $L_x/L_y$. 
This ratio increases with decreasing $L_x/L_y$ faster for the smaller boxes.
These measures of polymer position and organization provide a simple means
for quantitative characterization of the 2D probability distributions in 
Fig.~\ref{fig:prob2d_N60} that should be beneficial for comparison of results with those 
of future nanofluidics experiments employing rectangular cavities.
 
\begin{figure}[!ht]
\begin{center}
\vspace*{0.2in}
\includegraphics[angle=0,width=0.45\textwidth]{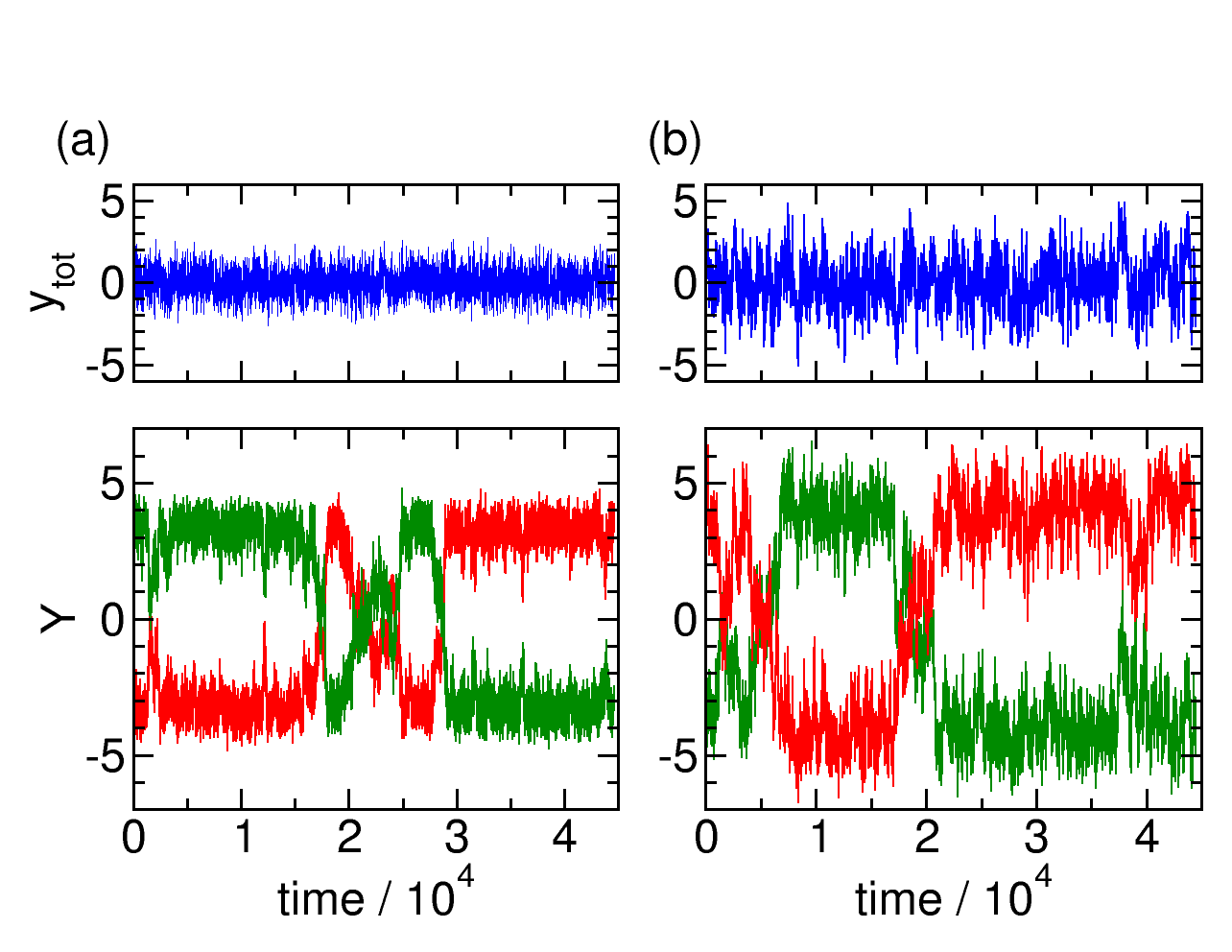}
\end{center}
\caption{Position histories for a system of two fully flexible LJ chains where 
$N$=60. (a) The top graph shows the history of combined COM of both polymers, 
$y_\text{tot}=Y_{1}+Y_{2}$, while the bottom graph shows the history of the coordinate 
of each polymer.  The rectangular cavity dimensions are such that $L_x/L_y=0.4$ and 
$\overline{L}/R_{{\rm g},xy}=1.58$.  (b) As  in (a), except for $L_x/L_y=0.5$ and 
$\overline{L}/R_{{\rm g},xy}=2.28$.} 
\label{fig:posyhist}
\end{figure}

Let us now consider the equilibrium dynamics of polymers confined to an anisometric box. 
Figure~\ref{fig:posyhist} shows {two} sample histories of {the
COM coordinate $Y$ for systems with polymers of length $N=60$ in cavities of different
size and shape.} Note that the centres of mass tend to be localized {at} 
positions at opposite ends of the elongated box, corresponding 
to the two high probability local regions in the ${\cal P}(x,y)$ distributions for elongated boxes
($L_x/L_y<1$) in Fig.~\ref{fig:prob2d_N60}. The histories also show that the polymers occasionally swap 
positions. The distributions of Fig.~\ref{fig:prob2d_N60} suggest that these swapping events will 
occur along a pathway that depends strongly on $L_x/L_y$. In the limit of large $L_x/L_y$, the 
polymers partially segregate along the $x$ axis as their centres cross the $y=0$ boundary.
By contrast, for more elongated boxes with small $L_x/L_y$, no such segregation along the $x$ 
dimension is observed, and the COM of each polymer passes through the centre of the box
during a chain swapping event. In this case, much greater interpenetration of the chains
occurs during the free energy barrier crossing.

\begin{figure}[!ht]
\begin{center}
\includegraphics[angle=0,width=0.45\textwidth]{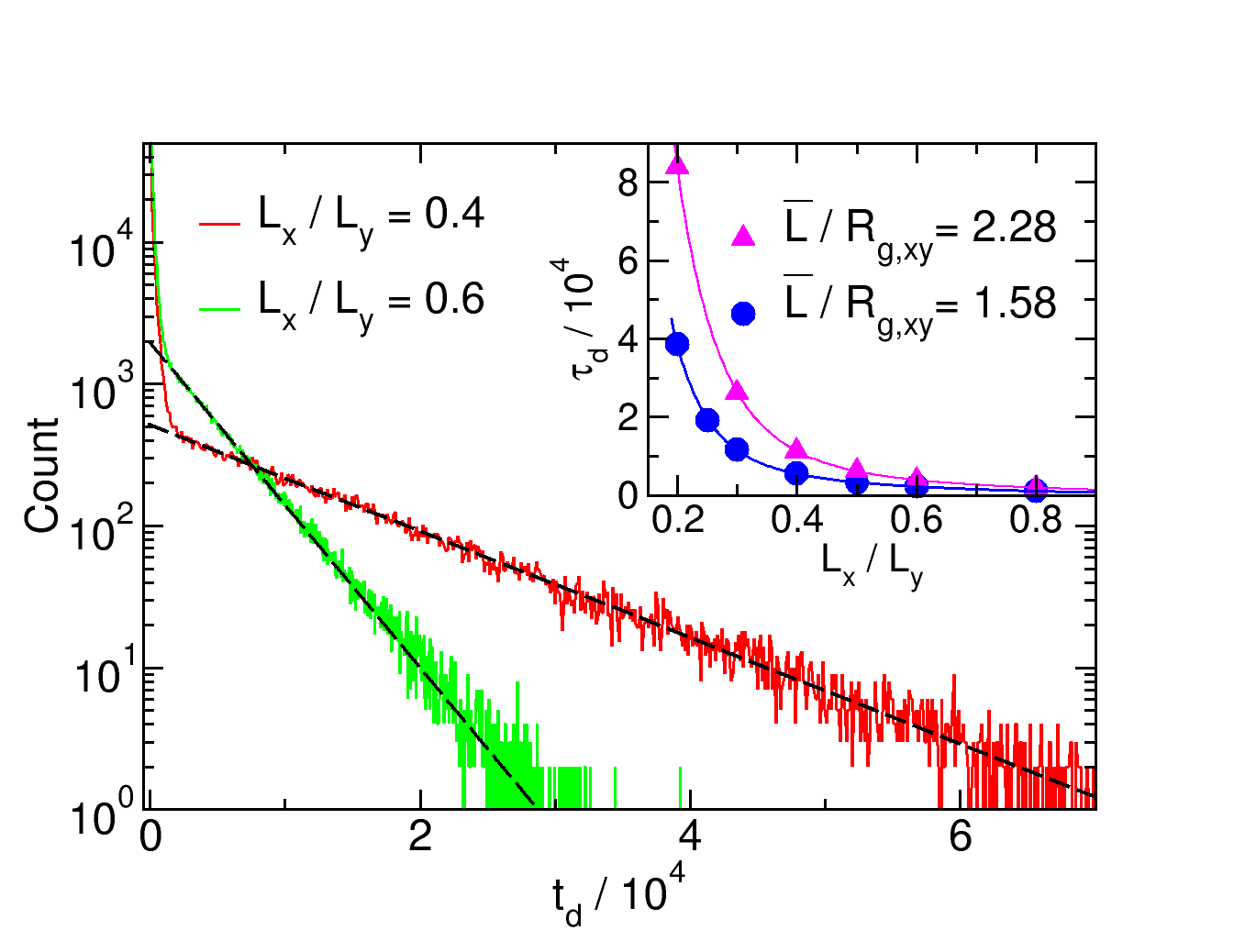}
\end{center}
\caption{Count distribution for the dwell times for a system with $N$=60 and
$\overline{L}/R_{{\rm g},xy}=2.28$. Results are shown for $L_x/L_y$=0.4 and $L_x/L_y$=0.6 for 
the red and green curves, respectively.  The dotted  lines are fits to a exponential 
of the form $\exp(-t_{\rm d}/\tau_{\rm d})$ with time constants of
$\tau_{\rm d}=(1.16\pm 0.01)\times 10^4$ for $L_x/L_y=4$ and 
$\tau_{\rm d}=(3.79\pm 0.03)\times 10^3$ for $L_x/L_y$=0.6. The inset shows the
variation of $\tau_{\rm d}$ with $L_x/L_y$ for two different scaled cavity sizes.}
\label{fig:SampleDwell}
\end{figure}

Let us consider the case of an elongated box with $L_x/L_y\lesssim 0.6$, where the 
the position swapping occurs via the latter mechanism described above. 
We define the dwell time $t_{\rm d}$ as the as the time between consecutive
crossings of the  $y$=0 boundary separating the two halves of the box. The
distribution of  the dwell times for two different box asymmetries is shown in
Fig.~\ref{fig:SampleDwell}. In each case, the distribution is exponential at 
long times with a notable deviation at very short times. The short-time probability
peak arises from rapid small-amplitude back-and-forth fluctuations as each
polymer passes through the box centre at a local free-energy maximum. A similar
feature was also noted in the experiments of Ref.~\onlinecite{liu2022confinement}.

The exponential form of the dwell-time distributions can be understood using
a simple model. First, note that polymer centres of mass tend to be strongly
anti-correlated along the $y$ axis, in the sense that if one polymer COM 
has a coordinate $y$, the other tends to have a coordinate of $-y$. Increasing
the box elongation tends to enhance this effect. Assuming (1) perfect
anticorrelation  between the COM positions of the polymers, (2) highly
localized centre-of-mass positions, and (3) that the time  spent transitioning
between states is negligible, the problem of tracking the dynamical behaviour 
of two polymers reduces to that of a single polymer that rapidly jumps
between two possible centre-of-mass positions. This is just a simple two-state
dynamical model. It is well known that the probability distribution for the time
spent in either state between successive jumps, $t_\text{d}$, for such a two-state
model described by a transition rate $k_0$ is given by\cite{Phillips_book}
\begin{eqnarray}
\mathcal{P}(t_\text{d})\propto \exp\left(-t_\text{d}/\tau_{\rm d}\right),
\end{eqnarray}
where the decay constant $\tau_{\rm d}$ is given by $\tau_{\rm d}=k_0^{-1}$. This
simple two-state model therefore accounts for the exponential decay of the
distributions shown in Fig.~\ref{fig:SampleDwell}. 

The values of the time constant $\tau_{\rm d}$ can easily be obtained by a fit
to the measured dwell time distributions. Sample fits are shown as dashed
lines in Fig.~\ref{fig:SampleDwell}. Note that we exclude the spurious peaks at
small dwell times from the fits. The inset shows the variation of the time
constant with respect to $L_x/L_y$ for two different values of the scaled
lateral box size, $\overline{L}/R_{{\rm g},xy}$. Generally, for fixed cross-sectional
area, $\tau_{\rm d}$ increases monotonically with increasing cavity elongation
(i.e., decreasing $L_x/L_y$ from unity). The results are qualitatively consistent
with comparable experimental results for elliptical cavities (see Fig.~3(d) of 
Ref.~\onlinecite{liu2022confinement}). The effect is diminished by reducing
the lateral cavity size, $\overline{L}/R_{{\rm g},xy}$.

The value of the time constant is largely determined by the features of the 
probability distributions in Fig.~\ref{fig:prob2d_N60}. This relationship can be 
elucidated using a multidimensional generalization of Kramers theory,\cite{matkowsky1988does}  
which predicts
\begin{eqnarray}
e^{\Delta F/k_{\rm B}T} = D\tau_\text{d}^*,
\label{eq:Kramer-Langer}
\end{eqnarray}
where
\begin{eqnarray}
\tau_{\rm d}^* \equiv \frac{Q_\text{B} \omega_\text{B} \omega_\text{W}}{2\pi Q_\text{W}} \tau_{\rm d}.
\end{eqnarray}
Here, $D$ is the Rouse diffusion coefficient, $\Delta F$ is the free energy barrier height,
and $\omega_\text{B}$ and $\omega_\text{W}$ represent the effective frequencies of the 
well and barrier, respectively.  In addition, $Q_\text{B}$ and $Q_\text{W}$ 
are the partition functions associated with the non-reactive modes at the free energy 
barrier and well, respectively. {The details for the procedure used
to extract $\omega_\text{B}$, $\omega_\text{W}$, $Q_\text{B}$ and $Q_\text{W}$ from
the free energy landscape are given in Section~I of the ESI.\dag  }

\begin{figure}[!ht]
\begin{center}
\includegraphics[angle=0,width=0.45\textwidth]{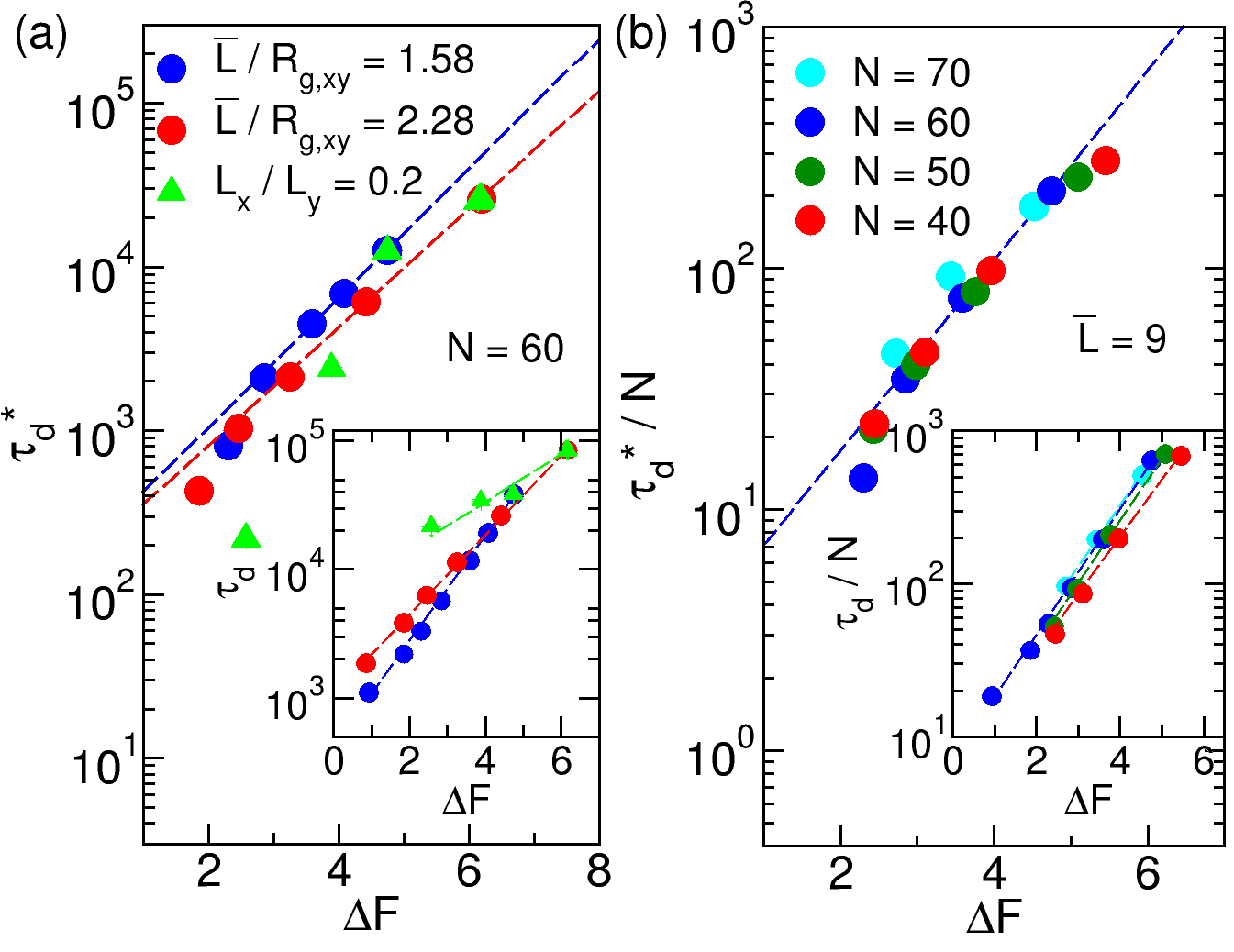}
\end{center}
\caption{(a) Scaled time constant, $\tau_{\rm d}^*$ (defined in Eq.~(\ref{eq:Kramer-Langer})) vs 
the free energy barrier height.  Circles indicate changing $\Delta F$ by varying 
$L_x/L_y$ for fixed $N$ and $\overline{L}$, and triangles indicate changing $\Delta F$
by varying $\overline{L}$ and fixing $L_x/L_y$ and $N$. The curves are fit to exponential 
functions, {$\tau_{\rm d}^* \propto e^{\alpha\Delta F}$},
with $\alpha=0.90\pm 0.03$ for $\overline{L}/R_{{\rm g},xy}=1.58$ and $\alpha=0.83\pm 0.02$ for 
$\overline{L}/R_{{\rm g},xy}=2.28$. The inset shows the unscaled time constant vs the barrier height. 
The {dashed} curves are {fits} to exponential functions.
(b) {$\tau_{\rm d}^*/N$ vs $\Delta F$ for several values of $N$ and for
$\overline{L}=9$.} The curves are fit to {the exponential function, 
$\tau_{\rm d}^*/N \propto e^{\alpha\Delta F}$,} with $\alpha$=0.90. 
The inset shows {$\tau_{\rm d}/N$} vs $\Delta F$. The
dashed lines are fits to exponential functions.
}
\label{fig:KramerTheory}
\end{figure}

Figure~\ref{fig:KramerTheory}(a) shows the variation of $\tau_{\rm d}^*$ with $\Delta F$.
Two data sets correspond to fixed box cross-sectional area, one with 
$\overline{L}/R_{{\rm g},xy}=1.58$ and the other with $\overline{L}/R_{{\rm g},xy}=2.28$. 
In each case, $\Delta F$ is controlled by variation in the box anisometry, $L_x/L_y$,
with larger anisometries leading to higher barriers. With the exception of the 
systems with very low $\Delta F$, where the Kramers theory is not expected to hold,
$\tau_{\rm d}^*$ {increases} exponentially with $\Delta F$, in agreement with the theoretical 
prediction of Eq.~(\ref{eq:Kramer-Langer}). The curves were fit to exponential functions 
of the form $\tau_\text{d}^*=C\exp(\alpha\Delta F)$.  In the case of 
$\overline{L}/R_{{\rm g},xy}=1.58$ we find $\alpha\approx 0.90\pm 0.03$, 
agreeing reasonably well with the predicted value of $\alpha=1$. However in 
the larger box size of $\overline{L}/R_{{\rm g},xy}=2.28$ we find $\alpha=0.83\pm 0.02$, 
which deviates a little more from the predicted value. In the
remaining data set, the anisometry is held fixed at $L_x/L_y=0.2$ and $\Delta F$ is 
controlled by varying the cross-sectional area. Here, $\tau_d^*$ does not appear to vary 
exponentially with $\Delta F$.  Figure~\ref{fig:KramerTheory}(b) shows the variation of 
$\tau_{\rm d}^*/N$ with $L_x/L_y$ for fixed
$\overline{L}$ for several values of $N$. Since the Rouse diffusion coefficient in
Eq.~(\ref{eq:Kramer-Langer}) scales as $D\propto 1/N$, the model predicts $\tau_d^*/N$ to be
invariant with respect to $N$. The data collapse onto a single curve is consistent with this
prediction. The curve is approximately exponential and characterized by $\alpha=0.9$, close
to the value of $\alpha=1$ predicted by the model.

\begin{figure}[!ht]
\begin{center}
\includegraphics[angle=0,width=0.45\textwidth]{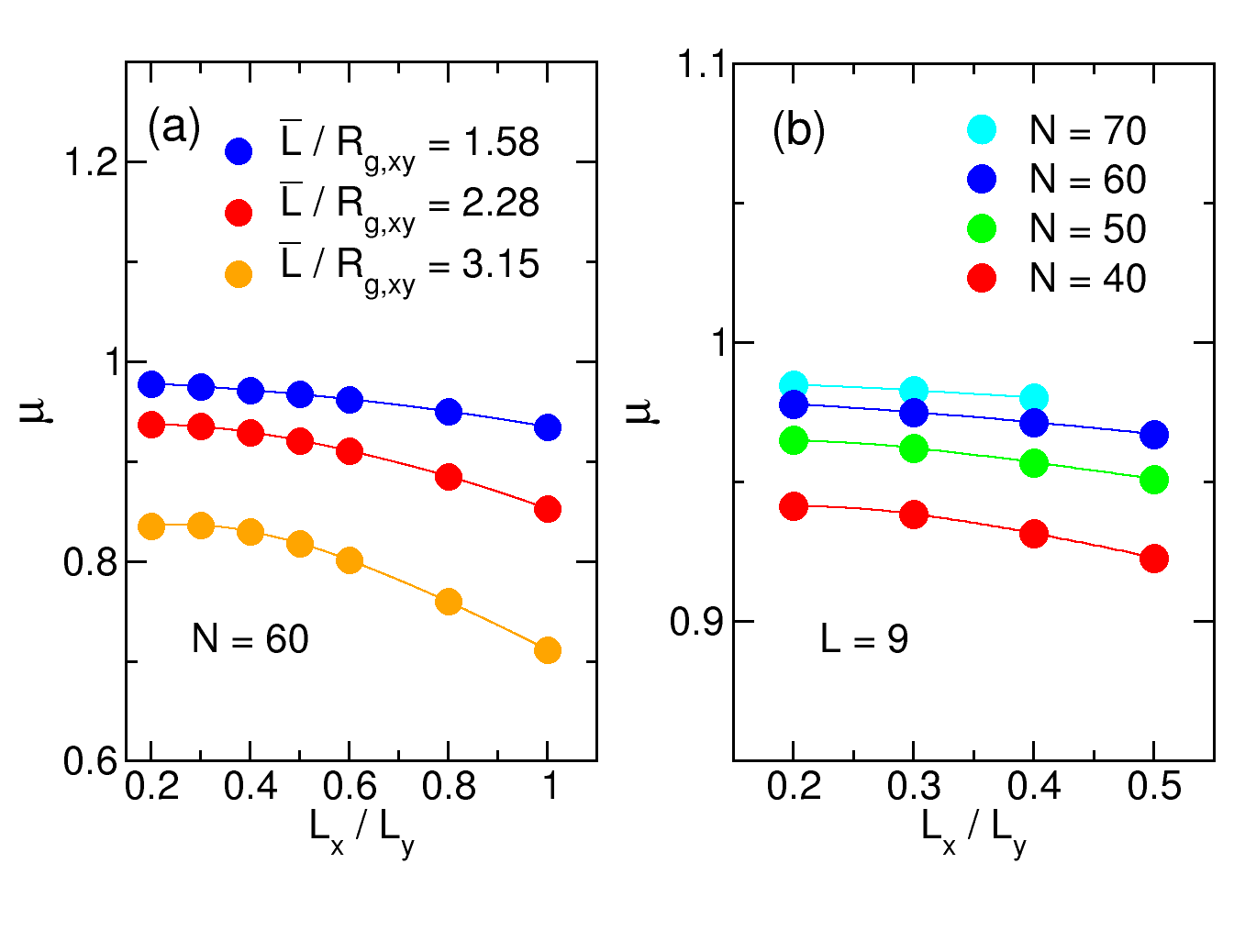}
\end{center}
\caption{COM position anticorrelation, $\mu$, for a system of two fully flexible LJ
polymers vs $L_x/L_y$ (a) for $N=60$ and several values of $\overline{L}/R_{{\rm g},xy}$,
and (b) for $\overline{L}/R_{{\rm g},xy}=1.58$ and several values of $N$.}
\label{fig:anticorr}
\end{figure}

As noted earlier, the theoretical model used to analyze the scaling of the dwell time constant
assumes perfect anticorrelation in the positions of the two polymers. The small deviation
between the observed and predicted scaling behaviour could arise in part from a breakdown
in this assumption.  To investigate this possibility, we measure the degree of anticorrelation, $\mu$,
between the COM {$y$-coordinates} of the two polymers, defined as
\begin{eqnarray}
\mu=-\frac{\langle Y_\text{1}Y_\text{2}\rangle}{\langle Y^2\rangle},
\label{eq:anti}
\end{eqnarray}
{where $\langle Y^2 \rangle \equiv \langle Y_1^2\rangle = \langle Y_2^2 \rangle$}.

Fig.~\ref{fig:anticorr}(a) shows the variation of $\mu$ with the box anisometry, $L_x/L_y$, for
three values of the scaled box size, $\overline{L}/R_{{\rm g},xy}$ for a chain length of $N=60$. 
The data correspond to those in Fig.~\ref{fig:KramerTheory}(a).  Two trends are apparent. 
First, position anticorrelation increases with decreasing box size. The data for the smallest 
box size of $\overline{L}/R_{{\rm g},xy}=1.58$ are closest to the value of $\mu=1$ 
{assumed in} the two-state model. In addition, for each cavity size, there is a small decrease 
in the position anticorrelation with decreasing cavity elongation (i.e., as $L_x/L_y$ increases),
with more a rapid decrease for the larger boxes.  The decrease in anticorrelation as both 
$\overline{L}$ and $L_x/L_y$ increase also correlates with the relative size of the fluctuations 
of {$y_\text{tot}=Y_1+Y_2$} in Fig.~\ref{fig:posyhist}. This arises from the fact that as $L_x/L_y$ 
decreases from unity the two probability peaks become more sharply localized at the positions on 
opposite sides of the cavity.  We speculate that the decrease in $\mu$ with $\overline{L}/R_{{\rm g},xy}$ 
is a primary cause for why the measured value of $\alpha$ obtained from fits to the data in 
Fig.~\ref{fig:KramerTheory}(a) decrease below the ideal value of unity with increasing
$\overline{L}/R_{{\rm g},xy}$.  It also likely accounts for the non-exponential behaviour 
in Fig.~\ref{fig:KramerTheory}(a) for the data set with fixed $L_x/L_y=0.2$ obtained by varying 
$\overline{L}/R_{{\rm g},xy}$.  Figure~\ref{fig:anticorr}(b) shows that the anticorrelation 
increases only slightly as the polymer length $N$ increases. Thus, varying $N$ over the range 
considered here is not expected to significantly affect the calculated value of $\alpha$, 
consistent with the data collapse observed in Fig.~\ref{fig:KramerTheory}(b).

{The effects of varying the cavity volume and elongation on the 
organization and dynamics of the polymers are qualitatively consistent with trends 
predicted by Jun and Wright using the blob model for a comparable physical 
system.\cite{jun2010entropy} In sufficiently elongated cavities, the blob size 
is comparable to the cavity width, and the strong repulsion between the blobs ensures
linear ordering along the cavity and, thus, chain segregation. For less elongated
cavities of the same volume, the blob size shrinks in size relative to the cavity
width, and inter-blob repulsion no longer enforces linear ordering. Instead, the 
string of blobs for each chain effectively undergoes a self-avoiding walk, resulting
in a mixed state in which the chains no longer segregate. These predictions were
corroborated by simulations.\cite{jung2012intrachain,jung2012ring} The increase
in chain miscibility with decreasing cavity elongation is accompanied by a
decrease in the free-energy barrier separating the most probable polymer 
centre-of-mass positions along the cavity, an effect quantified in 
Ref.~\onlinecite{polson2018segregation}. In both the theory and in these previous 
simulations, miscibility is enhanced by decreasing the cavity volume. All of these trends
are consistent with those evident in the probability maps of Fig.~\ref{fig:widthrectangle}
and \ref{fig:prob2d_N60}, and so account for the trends in the chain-swapping dynamics
in Figs.~\ref{fig:RMS-rectanglebox}, \ref{fig:posyhist}, \ref{fig:SampleDwell}, and 
\ref{fig:KramerTheory}. One feature of the present system that complicates direct
comparison with the predictions of Ref.~\onlinecite{jun2010entropy} is the particularly
small cavity dimension in the longitudinal direction ($h$), so chosen to mimic the 
experimental systems. This leads to confinement with three different dimensions for the 
elongated rectangular cavities, in contrast to the case in Ref.~\onlinecite{jun2010entropy},
where two cavity dimensions were equal.
In addition, the small system sizes employed here are likely too small for the blob
model to yield quantitatively accurate results. It is intriguing that the qualitative 
trends of the predictions nevertheless hold up. Further investigation using
much larger system sizes and varying $h$ the would be of value in assessing the
applicability of the theory to this system.}

To summarize, increasing cavity elongation in the lateral plane tends to promote segregation,
an effect that is enhanced by decreasing the lateral cavity size. A concomitant increase
in the free energy barrier between two localized positions
leads to a reduction in the rate of swapping. For sufficiently small cross-sectional area
and large cavity anisometry, the process is reasonably well described using a two-state
model together with a multi-dimensional extension of Kramers theory for activated processes. 
The observed small discrepancies arise from the marginal validity of some of the approximations,
notably the breakdown in the {assumption of perfect}  position anticorrelation of the 
two polymers.  The results presented here are in broad agreement with those of the recent experimental
study of Ref.~\onlinecite{liu2022confinement}. Note that the dynamics simulations employed
short ($N=60$ monomers) freely-jointed polymer chains due to the time-consuming nature
of the calculations. While this produces artificially wide polymers in relation to the $\lambda$ 
DNA chains used in the experiments the scaling analysis of Fig.~\ref{fig:widthrectangle} suggest 
comparable behaviour for realistic values of chain thickness $w$. 

\subsection{Elliptical cavities}
\label{subsec:elliptical}

The recent study by Liu {\it et al.}\cite{liu2022confinement} examined the behaviour
of two DNA molecules confined to a cavity with an elliptical cross section in the
lateral plane. They studied the configurational statistics and dynamics of two
different systems, one with two $\lambda$-DNA molecules and another consisting of a
T$_{\rm 4}$-DNA molecule and a single plasmid vector. Although
the two $\lambda$ DNA molecules have slightly different contour lengths due to
the use of different dyes, as noted earlier, they are of comparable magnitude.
On the other hand, the T$_{\rm 4}$-DNA molecule (166 kbp) is considerably longer
than the plasmid (4361 bp) and has a different topology (i.e. linear vs ring).
Not surprisingly, the two systems exhibit qualitatively different behaviour.
In this section, we carry out simulations for models that roughly correspond to
these different systems in order to better understand the trends observed in the
experiments. Note that the latter system exhibits asymmetries in both confinement
cavity dimensions and polymer length.

\begin{figure}[!ht]
\begin{center}
\vspace*{0.2in}
\includegraphics[angle=0,width=0.45\textwidth]{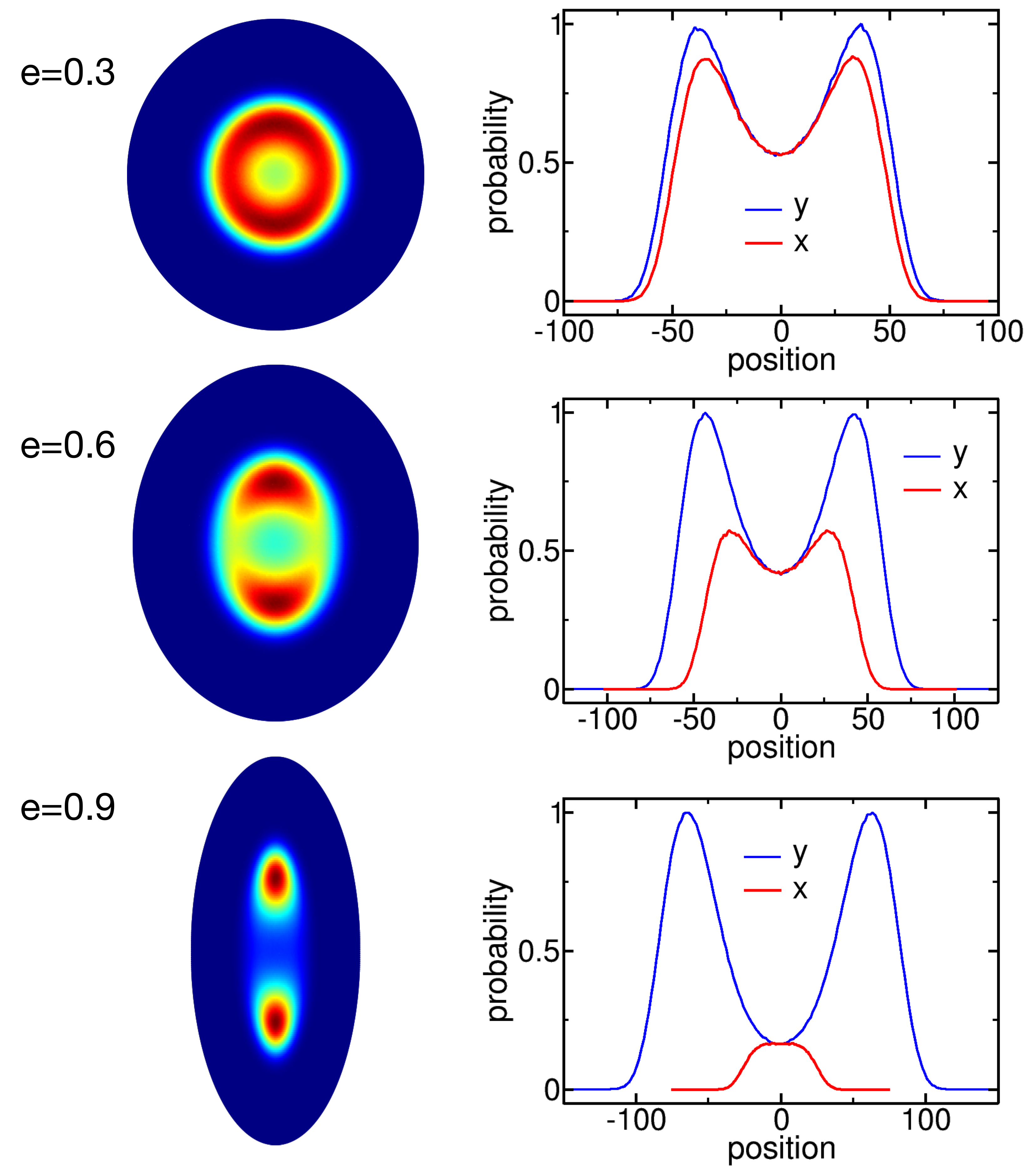}
\end{center}
\caption{COM probability distributions for a system of two semiflexible hard-sphere chains,
each of length $N=1900$ and bending rigidity $\kappa=6.36$, confined to a cavity
of height $h=20$ with an elliptical cross section of area $A=40000$. The left
column shows 2D distributions, ${\cal P}(x,y)$, for eccentricities of $e=0.3$, $0.6$
and $0.9$. The graphs in the right column show the corresponding cross sections of the
distributions along the vertical ($y$) and horizontal ($x$) symmetry axes of the distributions. }
\label{fig:prob_ellipse_N1900}
\end{figure}

In the first system, we use two identical semiflexible polymers, each of
length $N=1900$ monomers and bending rigidity $\kappa=6.36$. These values lead
to lengthscale ratios $\ell_{\rm c}/w$ and $P/w$ that correspond roughly to two
$\lambda$-DNA molecules stained with YOYO-1 under typical experimental conditions.
We use a cavity height of $h=20$ and a cross-sectional area of $A=40000$, 
corresponding to a ratio of $\bar{L}/R_{{\rm g},xy}=2.6$, where $\bar{L}\equiv
\sqrt{A}$ and the value of $R_{{\rm g},xy}$ is calculated for a
slit of height $h=20$. Figure~\ref{fig:prob_ellipse_N1900}
shows COM probability distributions, ${\cal P}(x,y)$, for confining cavities with
eccentricities of $e=0.3$, $0.6$ and $0.9$. The graphs on the right side of the
figure show probability cross sections through the vertical ($y$) and horizontal ($x$)
symmetry axes of the ellipses. The qualitative trends are similar to those observed
for rectangular boxes in Figs.~\ref{fig:widthrectangle} and \ref{fig:prob2d_N60}. 
The key trend is an increased tendency for the polymers to occupy two localized positions
as the eccentricity increases, i.e., as the box becomes more elongated. For a highly
elongated cavity with $e=0.9$, we note two quasi-discrete areas of  high probability 
along the long axis of the box,  indicating  the polymers segregate along the long 
axis. In this limit, the polymers are expected to undergo the same activated dynamics
described in the previous section. In the case of a low degree of cavity elongation 
for $e=0.3$,  the distribution has a ring-like structure with slight enhancements along 
the  long axis  of the box. This indicates that the polymers segregate; however, 
there is only a slight orientational preference due to the nearly circular shape
of the cavity. In this regime, the polymers are expected to undergo Brownian rotation.
In between these extremes at $e=0.6$, the distribution is somewhat more complex.
There is a faint probability ring as for more circular boxes, though the 
orientational preference is still relatively strong compared to $e$=0.3.
These results are qualitatively consistent with those for two $\lambda$-DNA molecules
confined to an elliptical cavity reported in Ref.~\onlinecite{liu2022confinement}.

\begin{figure*}[!ht]
\begin{center}
\vspace*{0.2in}
\includegraphics[angle=0,width=0.8\textwidth]{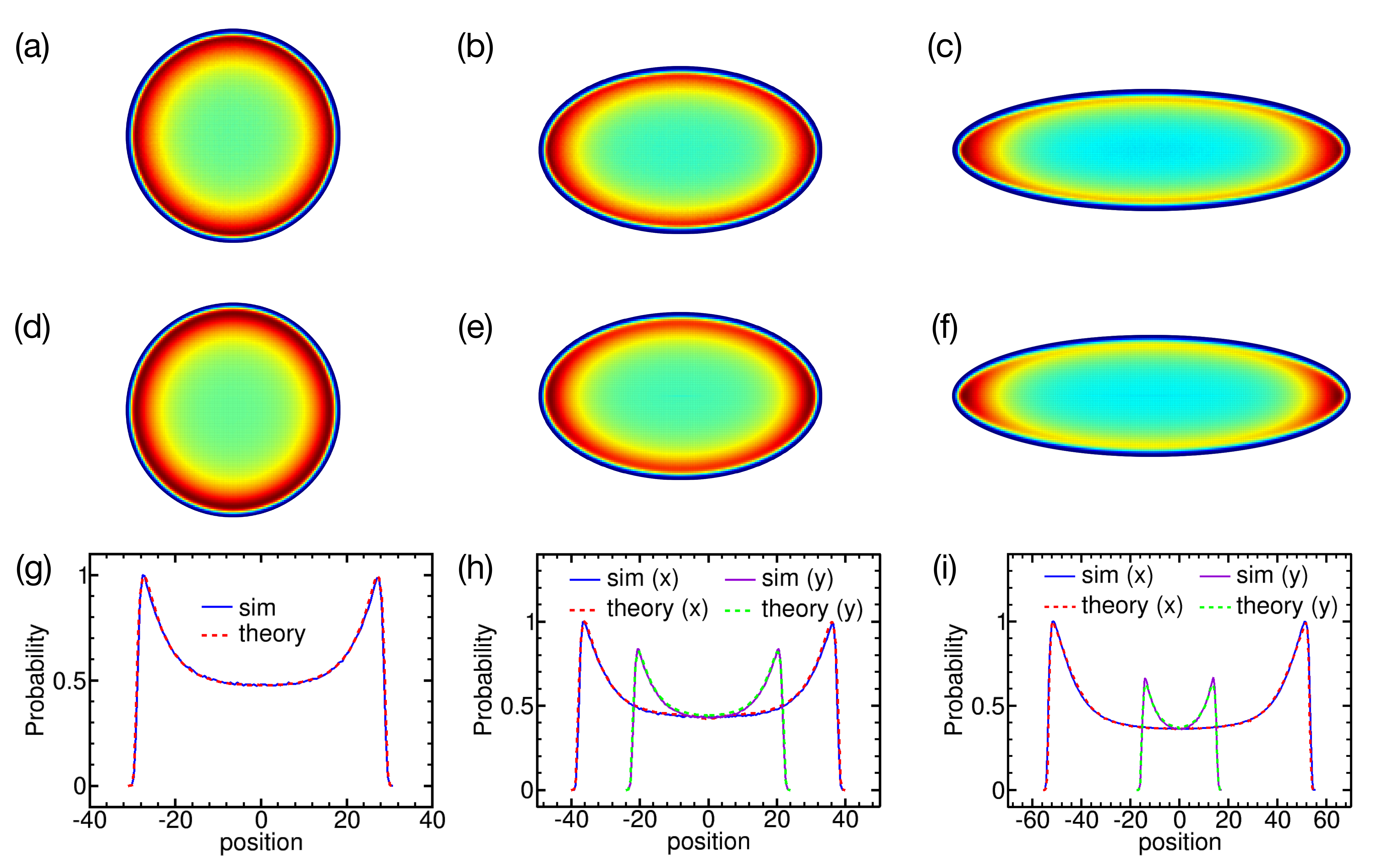}
\end{center}
\caption{COM probability, ${\cal P}(x,y)$, for a ring polymer confined with a large
linear polymer in a cavity with elliptical cross section. The cavity has a cross-sectional
area of $A=3000$ and a height of $h=15$. The ring polymer has a length of $N_2=25$,
and the linear polymer has a length of $N_1=1000$ and a bending rigidity of $\kappa=6.36$.
Panels (a), (b) and (c) show simulation results for cavities with eccentricity $e=0$,
$0.8$ and $0.95$, respectively. Panels (d), (e) and (f) show corresponding results from
a theoretical model for the distributions in the top row. The graphs in (g), (h) and (i)
show cross sections of the distributions along the horizontal ($x$) and vertical ($y$)
symmetry axes of the distributions.}
\label{fig:prob_theory_ellipse}
\end{figure*}

Next, we consider a two-polymer system consisting of a large linear polymer and a short
ring polymer. The long polymer is chosen to have a length of $N_1=1000$ monomers and a
bending rigidity of $\kappa=6.36$, while the ring polymer has a length of $N_2=25$. Note
that the contour-length ratio is comparable to that of the T$_{\rm 4}$/plasmid system
studied in Ref.~\onlinecite{liu2022confinement}. We choose a box with an elliptical
cross section of area $A=3000$ and a height of $h=15$. This corresponds to 
$\bar{L}/R_{{\rm g},xy}=0.94$, where $R_{{\rm g},xy}$ is the lateral rms radius
of gyration of the long polymer confined to a slit of $h=15$. This
is a very high degree of lateral confinement. 
Figure~\ref{fig:prob_theory_ellipse}(a)--(c) shows probability distributions of
the ring polymer for cavity eccentricities of $e=0$, $0.8$ and $0.95$. In all cases
there is an enhanced probability for the ring polymer to lie near the lateral wall
of the cavity. The origin of this trend is straightforward. This region corresponds
to an entropic depletion zone for the large polymer, and the short polymer is naturally
drawn into it to avoid collisions with the other polymer and thus increase its own
conformational entropy. As the cavity becomes more
elongated, the short polymer increasingly favours the regions near the poles of the 
ellipse over regions near the less curved parts of the wall. These trends are also
evident in the graphs in Figs.~\ref{fig:prob_theory_ellipse}(g)--(i), which show
cross sections of the distributions along the symmetry axes of the ellipses. The overall
trends are in qualitative agreement with the experimental results of 
Ref.~\onlinecite{liu2022confinement}. 
{Note that the ring topology of the short polymer has only a minor effect
on its position probability distribution. As evident in the results presented in section~II 
of the ESI\dag, employing linear topology for the short polymer increases only slightly the
entropic repulsion of the lateral confining walls, a consequence of the somewhat larger
average size of the linear polymer.  }

To better understand the observed bahaviour, we employ a theoretical model used in
Ref.~\onlinecite{liu2022confinement}, which was also used earlier in 
Section~\ref{subsec:AsymPA}, to understand polymer organization for polymers of
very different lengths confined to a square box (see Fig.~\ref{fig:prob_theory}).
As before, we construct a 2D free-energy function for the centre of mass of the
short (i.e., ring) polymer that is a sum of two contributions, 
$F(x,y) = F_{\rm int}(x,y) + F_{\rm wall}(x,y)$.
The interaction free energy with the large polymer is chosen to have the form
$F_{\rm int}(x,y) = a \rho_{\rm mon}(x,y)$, where $\rho_{\rm mon}(x,y)$ is
the monomer density of the large polymer, which is assumed to be unperturbed by the
presence of the short one. In addition, $F_{\rm wall}(x,y)$ is the free energy associated
with the interaction of the ring polymer with the lateral wall, which is only appreciable
when that polymer is very close to the wall. We calculate
$\rho_{\rm mon}(x,y)$ from a separate simulation with only the linear polymer
present. In addition, we calculate $F_{\rm wall}(x,y)$ using a multiple-histogram
MC simulation for a ring polymer of length $N_2=25$ confined to a slit of height $h=15$
and in the vicinity of a single hard flat wall in the lateral dimension. We find 
that the variation of $F_{\rm wall}$ with the COM distance from the wall is well
approximated by an exponential of the form $F_{\rm wall}(r) = b \exp(-r/r_b)$, where
$b=25.12$ and $r_b=1.60$. Calculation of the resulting probability distribution,
${\cal P}(x,y) \propto \exp(-F(x,y)/k_{\rm B}T)$ yields the results shown in panels 
(d), (e)  and (f) of Fig.~\ref{fig:prob_theory_ellipse} using values of $a/w^2=0.650$,
0.634 and 0.658 for $e=0.95$, $e=0.95$, and $e=0$, respectively. The 
calculated distributions match the corresponding simulation results very closely. A clearer
comparison is provided by the graphs shown in Figs.~\ref{fig:prob_theory_ellipse}(g)--(i),
in which cross sections of the calculated probability along the distribution 
symmetry axes are overlaid on those obtained directly from the simulation. 
These results are unambiguous: the theoretical model provides an exceptionally good 
prediction of the organization of the short ring polymer in the presence of a much
longer linear polymer confined to the elliptical cavity at sufficiently high density.

To assess the measured values of the coefficient, $a$, we follow 
the procedure in Ref.~\onlinecite{liu2022confinement}. The repulsive term in the Flory free
energy\cite{deGennes_book} is given by $k_{\rm B}T v_{\rm ex} \rho$, where $v_{\rm ex}$
is the excluded volume and $\rho$ is the Kuhn segment concentration.
Using the values of $\ell_{\rm k}$ and box height $h$ we convert $\rho$ to the
the 2D monomer density, $\rho_{\rm mon}$.  Next, we equate the repulsive
Flory free energy term to $F_{\rm int}=a\rho_{\rm mon}$ using a value of $a=0.65$
in order to calculate the effective excluded volume of the ring polymer. Approximating this
as a sphere of radius $r_{\rm p}$, we find that $r_{\rm p} \approx 3.1$, which is close
to the measured bulk RMS radius of gyration of the ring polymer of $R_{\rm g}=2.31$.
We conclude that {the} model for the interaction {between}
the short ring polymer {and} the linear polymer yields physically meaningful values of $a$.

In their analysis of the data for T$_{\rm 4}$-DNA/plasmid confinement in
elliptical cavities, Liu {\it et al.} found generally good agreement with the 
probability density cross sections. However, statistical scatter of the data precluded
a test of the model as quantitative as is provided in the present study. Their analysis
also involved estimation of the linear-polymer density profile by numerical solution of
a partial differential equation and invoking ground-state dominance. The authors note
that the calculated distribution is expected to become inaccurate near the poles
as the cavity becomes increasingly elongated. This is a regime where the semiflexiblity 
of the polymer is expected to become significant, a feature which was not incorporated
into the calculation. The analysis of the simulations show that modeling the interpolymer 
interactions with a free energy $F_{\rm int}\propto \rho_{\rm mon}(x,y)$ can indeed be a
decent approximation in the case where excluded-volume interactions dominate and provided
the polymer size ratio is sufficiently large. This should be useful information for
analysis of comparable data in future experiments. 
 
Of final note is the contrast in the quality of the theoretical predictions for the
elliptical cavities here and the square cavities in Sec.~\ref{subsec:AsymPA} 
{(see Fig.~\ref{fig:prob_theory})}. While
the predicted distributions in the latter case are qualitatively correct, the
quantitative discrepancies with the simulation data were significant. This is mostly
due to the larger size of the short polymer in that case, where the contour-length ratio 
was 5:1. By contrast, the size ratio here is close to 40:1, and the ring topology of the
small polymer contributes to making it even more compact. Both factors reduce the
degree to which the short polymer perturbs the configuration of the larger
one, which is an underlying assumption of the theoretical model. Finally, the magnitude
of the entropic repulsion of the short polymer with the wall in the corners of the 
square cavity is probably underestimated as a result of assuming the repulsion by
the two perpendicular walls is simply additive. This is what likely leads to the 
positioning of the  high-intensity peaks of Fig~\ref{fig:prob_theory}(a) closer to
the corners than is observed in the {simulations}. The lack of such sharp 
corners in the elliptical cavities together with using a much more compact short polymer 
precludes any such effect, except possibly in the limit of large $e$ where the wall 
curvature can be significant at the polar regions. We conclude that the theoretical model 
provides a reasonable prediction of the probability distribution for a sufficiently
compact polymer in the presence of a much larger one in the absence of sharp corners
or regions of {very} high curvature in the lateral confining wall.

\section{Conclusions}
\label{sec:conclusions}
In this study, we have employed both Brownian dynamics and Metropolis Monte Carlo 
simulations to investigate the organization and dynamics of two polymer chains confined 
to a box-like cavity with strong confinement in one-dimension. This work is motivated
by recent experimental studies of DNA molecules confined to nanocavities with both
square and elliptical cross sections. We examined the effects of varying polymer width, 
as well as asymmetries in polymer length and confinement cavity lateral dimensions.  
{We find that segregation is enhanced by increasing polymer width and 
that increasing cavity elongation promotes segregation and localization of identical 
polymers to opposite ends of the cavity. A free-energy barrier controls the rate of 
polymers swapping positions, and the observed dynamics are roughly consistent with
theoretical predictions and experimental results. Increasing the contour length 
difference between polymers significantly affects their organization in the cavity. 
In the case of a large linear polymer co-trapped with a small ring polymer in an 
elliptical cavity, the small polymer tends to lie near the lateral confining walls, 
and especially at the cavity poles for highly elongated ellipses.}

In future work, we will examine the effects of various other system features relevant to
recent and ongoing nanofluidics experiments of confined multi-chain systems, as well as 
{\it in vivo} experiments of chromosome segregation in prokaryotes. For example, in
Ref.~\onlinecite{liu2022confinement} small inert molecules of dextran were 
inserted into the cavity as a means to probe the effects of macromolecular crowding.  
The observed inward displacement of the plasmid probability density presumably arises 
solely from entropic effects, an assumption that can be tested via simulation.
Another variation of the system model is inspired by the behaviour of high copy 
number plasmids in {\it E. coli}, whose distribution tends to be remarkably
multi-focal in character, with large clusters near the cell poles,  despite lacking an 
active mechanism to ensure partitioning upon bacterial division.\cite{reyes2013high}
Simulations could examine {the} role and quantify the effect of chromosome-mediated 
entropic interactions between plasmids that may account for this effect.
More relevant for elucidating the mechanism of bacterial chromosome segregation is 
examining the effects of ring polymer topology as well as the effects of
loop formation through cross-linking, as in the phenomenon of proteins 
bridging chromosomal DNA segments.\cite{mitra2022polymer}
Above all, our simulation studies will continue to be guided by ongoing {\it in
vitro} DNA experiments employing nanofluidic devices.

\section*{Author contributions}

DR wrote the simulation code, carried out most of the simulations, analyzed the most
of the data, and wrote the first draft of the article. JP oversaw the project,
carried out some of the simulations and analysis and revised the manuscript.

\section*{Conflicts of interest}

There are no conflicts to declare.

\section*{Acknowledgements}
This work was supported by the Natural Sciences and Engineering Research Council
of Canada (NSERC). We are grateful to Compute Canada for use of their computational resources.
We would also like to thank W. Reisner and Z. Liu for helpful discussions.


\begin{thebibliography}{58}%
\makeatletter
\providecommand \@ifxundefined [1]{%
 \@ifx{#1\undefined}
}%
\providecommand \@ifnum [1]{%
 \ifnum #1\expandafter \@firstoftwo
 \else \expandafter \@secondoftwo
 \fi
}%
\providecommand \@ifx [1]{%
 \ifx #1\expandafter \@firstoftwo
 \else \expandafter \@secondoftwo
 \fi
}%
\providecommand \natexlab [1]{#1}%
\providecommand \enquote  [1]{``#1''}%
\providecommand \bibnamefont  [1]{#1}%
\providecommand \bibfnamefont [1]{#1}%
\providecommand \citenamefont [1]{#1}%
\providecommand \href@noop [0]{\@secondoftwo}%
\providecommand \href [0]{\begingroup \@sanitize@url \@href}%
\providecommand \@href[1]{\@@startlink{#1}\@@href}%
\providecommand \@@href[1]{\endgroup#1\@@endlink}%
\providecommand \@sanitize@url [0]{\catcode `\\12\catcode `\$12\catcode
  `\&12\catcode `\#12\catcode `\^12\catcode `\_12\catcode `\%12\relax}%
\providecommand \@@startlink[1]{}%
\providecommand \@@endlink[0]{}%
\providecommand \url  [0]{\begingroup\@sanitize@url \@url }%
\providecommand \@url [1]{\endgroup\@href {#1}{\urlprefix }}%
\providecommand \urlprefix  [0]{URL }%
\providecommand \Eprint [0]{\href }%
\providecommand \doibase [0]{http://dx.doi.org/}%
\providecommand \selectlanguage [0]{\@gobble}%
\providecommand \bibinfo  [0]{\@secondoftwo}%
\providecommand \bibfield  [0]{\@secondoftwo}%
\providecommand \translation [1]{[#1]}%
\providecommand \BibitemOpen [0]{}%
\providecommand \bibitemStop [0]{}%
\providecommand \bibitemNoStop [0]{.\EOS\space}%
\providecommand \EOS [0]{\spacefactor3000\relax}%
\providecommand \BibitemShut  [1]{\csname bibitem#1\endcsname}%
\let\auto@bib@innerbib\@empty
\bibitem [{\citenamefont {Jun}\ and\ \citenamefont
  {Mulder}(2006)}]{jun2006entropy}%
  \BibitemOpen
  \bibfield  {author} {\bibinfo {author} {\bibfnamefont {S.}~\bibnamefont
  {Jun}}\ and\ \bibinfo {author} {\bibfnamefont {B.}~\bibnamefont {Mulder}},\
  }\href@noop {} {\bibfield  {journal} {\bibinfo  {journal} {{ Proc. Natl.
  Acad. Sci. USA}}\ }\textbf {\bibinfo {volume} {103}},\ \bibinfo {pages}
  {12388} (\bibinfo {year} {2006})}\BibitemShut {NoStop}%
\bibitem [{\citenamefont {Teraoka}\ and\ \citenamefont
  {Wang}(2004)}]{teraoka2004computer}%
  \BibitemOpen
  \bibfield  {author} {\bibinfo {author} {\bibfnamefont {I.}~\bibnamefont
  {Teraoka}}\ and\ \bibinfo {author} {\bibfnamefont {Y.}~\bibnamefont {Wang}},\
  }\href@noop {} {\bibfield  {journal} {\bibinfo  {journal} {Polymer}\ }\textbf
  {\bibinfo {volume} {45}},\ \bibinfo {pages} {3835} (\bibinfo {year}
  {2004})}\BibitemShut {NoStop}%
\bibitem [{\citenamefont {Jun}\ \emph {et~al.}(2007)\citenamefont {Jun},
  \citenamefont {Arnold},\ and\ \citenamefont {Ha}}]{jun2007confined}%
  \BibitemOpen
  \bibfield  {author} {\bibinfo {author} {\bibfnamefont {S.}~\bibnamefont
  {Jun}}, \bibinfo {author} {\bibfnamefont {A.}~\bibnamefont {Arnold}}, \ and\
  \bibinfo {author} {\bibfnamefont {B.-Y.}\ \bibnamefont {Ha}},\ }\href@noop {}
  {\bibfield  {journal} {\bibinfo  {journal} {Phys. Rev. Lett.}\ }\textbf
  {\bibinfo {volume} {98}},\ \bibinfo {pages} {128303} (\bibinfo {year}
  {2007})}\BibitemShut {NoStop}%
\bibitem [{\citenamefont {Arnold}\ and\ \citenamefont
  {Jun}(2007)}]{arnold2007time}%
  \BibitemOpen
  \bibfield  {author} {\bibinfo {author} {\bibfnamefont {A.}~\bibnamefont
  {Arnold}}\ and\ \bibinfo {author} {\bibfnamefont {S.}~\bibnamefont {Jun}},\
  }\href@noop {} {\bibfield  {journal} {\bibinfo  {journal} {Phys. Rev. E}\
  }\textbf {\bibinfo {volume} {76}},\ \bibinfo {pages} {031901} (\bibinfo
  {year} {2007})}\BibitemShut {NoStop}%
\bibitem [{\citenamefont {Jacobsen}(2010)}]{jacobsen2010demixing}%
  \BibitemOpen
  \bibfield  {author} {\bibinfo {author} {\bibfnamefont {J.~L.}\ \bibnamefont
  {Jacobsen}},\ }\href@noop {} {\bibfield  {journal} {\bibinfo  {journal}
  {Phys. Rev. E}\ }\textbf {\bibinfo {volume} {82}},\ \bibinfo {pages} {051802}
  (\bibinfo {year} {2010})}\BibitemShut {NoStop}%
\bibitem [{\citenamefont {Jung}\ and\ \citenamefont
  {Ha}(2010)}]{jung2010overlapping}%
  \BibitemOpen
  \bibfield  {author} {\bibinfo {author} {\bibfnamefont {Y.}~\bibnamefont
  {Jung}}\ and\ \bibinfo {author} {\bibfnamefont {B.-Y.}\ \bibnamefont {Ha}},\
  }\href@noop {} {\bibfield  {journal} {\bibinfo  {journal} {Phys. Rev. E}\
  }\textbf {\bibinfo {volume} {82}},\ \bibinfo {pages} {051926} (\bibinfo
  {year} {2010})}\BibitemShut {NoStop}%
\bibitem [{\citenamefont {Jung}\ \emph
  {et~al.}(2012{\natexlab{a}})\citenamefont {Jung}, \citenamefont {Jeon},
  \citenamefont {Kim}, \citenamefont {Jeong}, \citenamefont {Jun},\ and\
  \citenamefont {Ha}}]{jung2012ring}%
  \BibitemOpen
  \bibfield  {author} {\bibinfo {author} {\bibfnamefont {Y.}~\bibnamefont
  {Jung}}, \bibinfo {author} {\bibfnamefont {C.}~\bibnamefont {Jeon}}, \bibinfo
  {author} {\bibfnamefont {J.}~\bibnamefont {Kim}}, \bibinfo {author}
  {\bibfnamefont {H.}~\bibnamefont {Jeong}}, \bibinfo {author} {\bibfnamefont
  {S.}~\bibnamefont {Jun}}, \ and\ \bibinfo {author} {\bibfnamefont {B.-Y.}\
  \bibnamefont {Ha}},\ }\href@noop {} {\bibfield  {journal} {\bibinfo
  {journal} {Soft Matter}\ }\textbf {\bibinfo {volume} {8}},\ \bibinfo {pages}
  {2095} (\bibinfo {year} {2012}{\natexlab{a}})}\BibitemShut {NoStop}%
\bibitem [{\citenamefont {Jung}\ \emph
  {et~al.}(2012{\natexlab{b}})\citenamefont {Jung}, \citenamefont {Kim},
  \citenamefont {Jun},\ and\ \citenamefont {Ha}}]{jung2012intrachain}%
  \BibitemOpen
  \bibfield  {author} {\bibinfo {author} {\bibfnamefont {Y.}~\bibnamefont
  {Jung}}, \bibinfo {author} {\bibfnamefont {J.}~\bibnamefont {Kim}}, \bibinfo
  {author} {\bibfnamefont {S.}~\bibnamefont {Jun}}, \ and\ \bibinfo {author}
  {\bibfnamefont {B.-Y.}\ \bibnamefont {Ha}},\ }\href@noop {} {\bibfield
  {journal} {\bibinfo  {journal} {Macromolecules}\ }\textbf {\bibinfo {volume}
  {45}},\ \bibinfo {pages} {3256} (\bibinfo {year}
  {2012}{\natexlab{b}})}\BibitemShut {NoStop}%
\bibitem [{\citenamefont {Liu}\ and\ \citenamefont
  {Chakraborty}(2012)}]{liu2012segregation}%
  \BibitemOpen
  \bibfield  {author} {\bibinfo {author} {\bibfnamefont {Y.}~\bibnamefont
  {Liu}}\ and\ \bibinfo {author} {\bibfnamefont {B.}~\bibnamefont
  {Chakraborty}},\ }\href@noop {} {\bibfield  {journal} {\bibinfo  {journal}
  {Phys. Biol.}\ }\textbf {\bibinfo {volume} {9}},\ \bibinfo {pages} {066005}
  (\bibinfo {year} {2012})}\BibitemShut {NoStop}%
\bibitem [{\citenamefont {Dorier}\ and\ \citenamefont
  {Stasiak}(2013)}]{dorier2013modelling}%
  \BibitemOpen
  \bibfield  {author} {\bibinfo {author} {\bibfnamefont {J.}~\bibnamefont
  {Dorier}}\ and\ \bibinfo {author} {\bibfnamefont {A.}~\bibnamefont
  {Stasiak}},\ }\href@noop {} {\bibfield  {journal} {\bibinfo  {journal}
  {Nucleic Acids Res.}\ }\textbf {\bibinfo {volume} {41}},\ \bibinfo {pages}
  {6808} (\bibinfo {year} {2013})}\BibitemShut {NoStop}%
\bibitem [{\citenamefont {Ra{\v{c}}ko}\ and\ \citenamefont
  {Cifra}(2013)}]{racko2013segregation}%
  \BibitemOpen
  \bibfield  {author} {\bibinfo {author} {\bibfnamefont {D.}~\bibnamefont
  {Ra{\v{c}}ko}}\ and\ \bibinfo {author} {\bibfnamefont {P.}~\bibnamefont
  {Cifra}},\ }\href@noop {} {\bibfield  {journal} {\bibinfo  {journal} {J.
  Chem. Phys.}\ }\textbf {\bibinfo {volume} {138}},\ \bibinfo {pages} {184904}
  (\bibinfo {year} {2013})}\BibitemShut {NoStop}%
\bibitem [{\citenamefont {Shin}\ \emph {et~al.}(2014)\citenamefont {Shin},
  \citenamefont {Cherstvy},\ and\ \citenamefont {Metzler}}]{shin2014mixing}%
  \BibitemOpen
  \bibfield  {author} {\bibinfo {author} {\bibfnamefont {J.}~\bibnamefont
  {Shin}}, \bibinfo {author} {\bibfnamefont {A.~G.}\ \bibnamefont {Cherstvy}},
  \ and\ \bibinfo {author} {\bibfnamefont {R.}~\bibnamefont {Metzler}},\
  }\href@noop {} {\bibfield  {journal} {\bibinfo  {journal} {New J. Phys.}\
  }\textbf {\bibinfo {volume} {16}},\ \bibinfo {pages} {053047} (\bibinfo
  {year} {2014})}\BibitemShut {NoStop}%
\bibitem [{\citenamefont {Minina}\ and\ \citenamefont
  {Arnold}(2014)}]{minina2014induction}%
  \BibitemOpen
  \bibfield  {author} {\bibinfo {author} {\bibfnamefont {E.}~\bibnamefont
  {Minina}}\ and\ \bibinfo {author} {\bibfnamefont {A.}~\bibnamefont
  {Arnold}},\ }\href@noop {} {\bibfield  {journal} {\bibinfo  {journal} {Soft
  Matter}\ }\textbf {\bibinfo {volume} {10}},\ \bibinfo {pages} {5836}
  (\bibinfo {year} {2014})}\BibitemShut {NoStop}%
\bibitem [{\citenamefont {Minina}\ and\ \citenamefont
  {Arnold}(2015)}]{minina2015entropic}%
  \BibitemOpen
  \bibfield  {author} {\bibinfo {author} {\bibfnamefont {E.}~\bibnamefont
  {Minina}}\ and\ \bibinfo {author} {\bibfnamefont {A.}~\bibnamefont
  {Arnold}},\ }\href@noop {} {\bibfield  {journal} {\bibinfo  {journal}
  {Macromolecules}\ }\textbf {\bibinfo {volume} {48}},\ \bibinfo {pages} {4998}
  (\bibinfo {year} {2015})}\BibitemShut {NoStop}%
\bibitem [{\citenamefont {Chen}\ \emph {et~al.}(2015)\citenamefont {Chen},
  \citenamefont {Yu}, \citenamefont {Wang},\ and\ \citenamefont
  {Luo}}]{chen2015polymer}%
  \BibitemOpen
  \bibfield  {author} {\bibinfo {author} {\bibfnamefont {Y.}~\bibnamefont
  {Chen}}, \bibinfo {author} {\bibfnamefont {W.}~\bibnamefont {Yu}}, \bibinfo
  {author} {\bibfnamefont {J.}~\bibnamefont {Wang}}, \ and\ \bibinfo {author}
  {\bibfnamefont {K.}~\bibnamefont {Luo}},\ }\href@noop {} {\bibfield
  {journal} {\bibinfo  {journal} {J. Chem. Phys.}\ }\textbf {\bibinfo {volume}
  {143}},\ \bibinfo {pages} {134904} (\bibinfo {year} {2015})}\BibitemShut
  {NoStop}%
\bibitem [{\citenamefont {Polson}\ and\ \citenamefont
  {Montgomery}(2014)}]{polson2014polymer}%
  \BibitemOpen
  \bibfield  {author} {\bibinfo {author} {\bibfnamefont {J.~M.}\ \bibnamefont
  {Polson}}\ and\ \bibinfo {author} {\bibfnamefont {L.~G.}\ \bibnamefont
  {Montgomery}},\ }\href@noop {} {\bibfield  {journal} {\bibinfo  {journal} {J.
  Chem. Phys.}\ }\textbf {\bibinfo {volume} {141}},\ \bibinfo {pages} {164902}
  (\bibinfo {year} {2014})}\BibitemShut {NoStop}%
\bibitem [{\citenamefont {Du}\ \emph {et~al.}(2018)\citenamefont {Du},
  \citenamefont {Jiang},\ and\ \citenamefont {Hou}}]{du2018polymer}%
  \BibitemOpen
  \bibfield  {author} {\bibinfo {author} {\bibfnamefont {Y.}~\bibnamefont
  {Du}}, \bibinfo {author} {\bibfnamefont {H.}~\bibnamefont {Jiang}}, \ and\
  \bibinfo {author} {\bibfnamefont {Z.}~\bibnamefont {Hou}},\ }\href@noop {}
  {\bibfield  {journal} {\bibinfo  {journal} {J. Chem. Phys.}\ }\textbf
  {\bibinfo {volume} {149}},\ \bibinfo {pages} {244906} (\bibinfo {year}
  {2018})}\BibitemShut {NoStop}%
\bibitem [{\citenamefont {Polson}\ and\ \citenamefont
  {Kerry}(2018)}]{polson2018segregation}%
  \BibitemOpen
  \bibfield  {author} {\bibinfo {author} {\bibfnamefont {J.~M.}\ \bibnamefont
  {Polson}}\ and\ \bibinfo {author} {\bibfnamefont {D.~R.-M.}\ \bibnamefont
  {Kerry}},\ }\href@noop {} {\bibfield  {journal} {\bibinfo  {journal} {Soft
  Matter}\ }\textbf {\bibinfo {volume} {14}},\ \bibinfo {pages} {6360}
  (\bibinfo {year} {2018})}\BibitemShut {NoStop}%
\bibitem [{\citenamefont
  {Nowicki}(2019{\natexlab{a}})}]{nowicki2019segregation}%
  \BibitemOpen
  \bibfield  {author} {\bibinfo {author} {\bibfnamefont {W.}~\bibnamefont
  {Nowicki}},\ }\href@noop {} {\bibfield  {journal} {\bibinfo  {journal} {J.
  Chem. Phys.}\ }\textbf {\bibinfo {volume} {150}},\ \bibinfo {pages} {014902}
  (\bibinfo {year} {2019}{\natexlab{a}})}\BibitemShut {NoStop}%
\bibitem [{\citenamefont
  {Nowicki}(2019{\natexlab{b}})}]{nowicki2019electrostatic}%
  \BibitemOpen
  \bibfield  {author} {\bibinfo {author} {\bibfnamefont {W.}~\bibnamefont
  {Nowicki}},\ }\href@noop {} {\bibfield  {journal} {\bibinfo  {journal} {J.
  Mol. Model.}\ }\textbf {\bibinfo {volume} {25}},\ \bibinfo {pages} {269}
  (\bibinfo {year} {2019}{\natexlab{b}})}\BibitemShut {NoStop}%
\bibitem [{\citenamefont {Polson}\ and\ \citenamefont
  {Zhu}(2021)}]{polson2021free}%
  \BibitemOpen
  \bibfield  {author} {\bibinfo {author} {\bibfnamefont {J.~M.}\ \bibnamefont
  {Polson}}\ and\ \bibinfo {author} {\bibfnamefont {Q.}~\bibnamefont {Zhu}},\
  }\href {\doibase 10.1103/PhysRevE.103.012501} {\bibfield  {journal} {\bibinfo
   {journal} {Phys. Rev. E}\ }\textbf {\bibinfo {volume} {103}},\ \bibinfo
  {pages} {012501} (\bibinfo {year} {2021})}\BibitemShut {NoStop}%
\bibitem [{\citenamefont {Mitra}\ \emph
  {et~al.}(2022{\natexlab{a}})\citenamefont {Mitra}, \citenamefont {Pande},\
  and\ \citenamefont {Chatterji}}]{mitra2022polymer}%
  \BibitemOpen
  \bibfield  {author} {\bibinfo {author} {\bibfnamefont {D.}~\bibnamefont
  {Mitra}}, \bibinfo {author} {\bibfnamefont {S.}~\bibnamefont {Pande}}, \ and\
  \bibinfo {author} {\bibfnamefont {A.}~\bibnamefont {Chatterji}},\ }\href@noop
  {} {\bibfield  {journal} {\bibinfo  {journal} {Soft Matter}\ }\textbf
  {\bibinfo {volume} {18}},\ \bibinfo {pages} {5615} (\bibinfo {year}
  {2022}{\natexlab{a}})}\BibitemShut {NoStop}%
\bibitem [{\citenamefont {Mitra}\ \emph
  {et~al.}(2022{\natexlab{b}})\citenamefont {Mitra}, \citenamefont {Pande},\
  and\ \citenamefont {Chatterji}}]{mitra2022topology}%
  \BibitemOpen
  \bibfield  {author} {\bibinfo {author} {\bibfnamefont {D.}~\bibnamefont
  {Mitra}}, \bibinfo {author} {\bibfnamefont {S.}~\bibnamefont {Pande}}, \ and\
  \bibinfo {author} {\bibfnamefont {A.}~\bibnamefont {Chatterji}},\ }\href@noop
  {} {\bibfield  {journal} {\bibinfo  {journal} {Physical Review E}\ }\textbf
  {\bibinfo {volume} {106}},\ \bibinfo {pages} {054502} (\bibinfo {year}
  {2022}{\natexlab{b}})}\BibitemShut {NoStop}%
\bibitem [{\citenamefont {Dorfman}\ \emph {et~al.}(2012)\citenamefont
  {Dorfman}, \citenamefont {King}, \citenamefont {Olson}, \citenamefont
  {Thomas},\ and\ \citenamefont {Tree}}]{dorfman2012beyond}%
  \BibitemOpen
  \bibfield  {author} {\bibinfo {author} {\bibfnamefont {K.~D.}\ \bibnamefont
  {Dorfman}}, \bibinfo {author} {\bibfnamefont {S.~B.}\ \bibnamefont {King}},
  \bibinfo {author} {\bibfnamefont {D.~W.}\ \bibnamefont {Olson}}, \bibinfo
  {author} {\bibfnamefont {J.~D.}\ \bibnamefont {Thomas}}, \ and\ \bibinfo
  {author} {\bibfnamefont {D.~R.}\ \bibnamefont {Tree}},\ }\href@noop {}
  {\bibfield  {journal} {\bibinfo  {journal} {Chem. Rev.}\ }\textbf {\bibinfo
  {volume} {113}},\ \bibinfo {pages} {2584} (\bibinfo {year}
  {2012})}\BibitemShut {NoStop}%
\bibitem [{\citenamefont {Reisner}\ \emph {et~al.}(2010)\citenamefont
  {Reisner}, \citenamefont {Larsen}, \citenamefont {Silahtaroglu},
  \citenamefont {Kristensen}, \citenamefont {Tommerup}, \citenamefont
  {Tegenfeldt},\ and\ \citenamefont {Flyvbjerg}}]{reisner2010single}%
  \BibitemOpen
  \bibfield  {author} {\bibinfo {author} {\bibfnamefont {W.}~\bibnamefont
  {Reisner}}, \bibinfo {author} {\bibfnamefont {N.~B.}\ \bibnamefont {Larsen}},
  \bibinfo {author} {\bibfnamefont {A.}~\bibnamefont {Silahtaroglu}}, \bibinfo
  {author} {\bibfnamefont {A.}~\bibnamefont {Kristensen}}, \bibinfo {author}
  {\bibfnamefont {N.}~\bibnamefont {Tommerup}}, \bibinfo {author}
  {\bibfnamefont {J.~O.}\ \bibnamefont {Tegenfeldt}}, \ and\ \bibinfo {author}
  {\bibfnamefont {H.}~\bibnamefont {Flyvbjerg}},\ }\href@noop {} {\bibfield
  {journal} {\bibinfo  {journal} {Proc. Natl. Acad. Sci. USA}\ }\textbf
  {\bibinfo {volume} {107}},\ \bibinfo {pages} {13294} (\bibinfo {year}
  {2010})}\BibitemShut {NoStop}%
\bibitem [{\citenamefont {Marie}\ \emph {et~al.}(2013)\citenamefont {Marie},
  \citenamefont {Pedersen}, \citenamefont {Bauer}, \citenamefont {Rasmussen},
  \citenamefont {Yusuf}, \citenamefont {Volpi}, \citenamefont {Flyvbjerg},
  \citenamefont {Kristensen},\ and\ \citenamefont {Mir}}]{marie2013integrated}%
  \BibitemOpen
  \bibfield  {author} {\bibinfo {author} {\bibfnamefont {R.}~\bibnamefont
  {Marie}}, \bibinfo {author} {\bibfnamefont {J.~N.}\ \bibnamefont {Pedersen}},
  \bibinfo {author} {\bibfnamefont {D.~L.}\ \bibnamefont {Bauer}}, \bibinfo
  {author} {\bibfnamefont {K.~H.}\ \bibnamefont {Rasmussen}}, \bibinfo {author}
  {\bibfnamefont {M.}~\bibnamefont {Yusuf}}, \bibinfo {author} {\bibfnamefont
  {E.}~\bibnamefont {Volpi}}, \bibinfo {author} {\bibfnamefont
  {H.}~\bibnamefont {Flyvbjerg}}, \bibinfo {author} {\bibfnamefont
  {A.}~\bibnamefont {Kristensen}}, \ and\ \bibinfo {author} {\bibfnamefont
  {K.~U.}\ \bibnamefont {Mir}},\ }\href@noop {} {\bibfield  {journal} {\bibinfo
   {journal} {Proc. Natl. Acad. Sci. USA}\ }\textbf {\bibinfo {volume} {110}},\
  \bibinfo {pages} {4893} (\bibinfo {year} {2013})}\BibitemShut {NoStop}%
\bibitem [{\citenamefont {Lam}\ \emph {et~al.}(2012)\citenamefont {Lam},
  \citenamefont {Hastie}, \citenamefont {Lin}, \citenamefont {Ehrlich},
  \citenamefont {Das}, \citenamefont {Austin}, \citenamefont {Deshpande},
  \citenamefont {Cao}, \citenamefont {Nagarajan}, \citenamefont {Xiao},\ and\
  \citenamefont {Kwok}}]{lam2012genome}%
  \BibitemOpen
  \bibfield  {author} {\bibinfo {author} {\bibfnamefont {E.~T.}\ \bibnamefont
  {Lam}}, \bibinfo {author} {\bibfnamefont {A.}~\bibnamefont {Hastie}},
  \bibinfo {author} {\bibfnamefont {C.}~\bibnamefont {Lin}}, \bibinfo {author}
  {\bibfnamefont {D.}~\bibnamefont {Ehrlich}}, \bibinfo {author} {\bibfnamefont
  {S.~K.}\ \bibnamefont {Das}}, \bibinfo {author} {\bibfnamefont {M.~D.}\
  \bibnamefont {Austin}}, \bibinfo {author} {\bibfnamefont {P.}~\bibnamefont
  {Deshpande}}, \bibinfo {author} {\bibfnamefont {H.}~\bibnamefont {Cao}},
  \bibinfo {author} {\bibfnamefont {N.}~\bibnamefont {Nagarajan}}, \bibinfo
  {author} {\bibfnamefont {M.}~\bibnamefont {Xiao}}, \ and\ \bibinfo {author}
  {\bibfnamefont {P.-Y.}\ \bibnamefont {Kwok}},\ }\href@noop {} {\bibfield
  {journal} {\bibinfo  {journal} {Nat. Biotech.}\ }\textbf {\bibinfo {volume}
  {30}},\ \bibinfo {pages} {771} (\bibinfo {year} {2012})}\BibitemShut
  {NoStop}%
\bibitem [{\citenamefont {Hastie}\ \emph {et~al.}(2013)\citenamefont {Hastie},
  \citenamefont {Dong}, \citenamefont {Smith}, \citenamefont {Finklestein},
  \citenamefont {Lam}, \citenamefont {Huo}, \citenamefont {Cao}, \citenamefont
  {Kwok}, \citenamefont {Deal},\ and\ \citenamefont
  {Dvorak}}]{hastie2013rapid}%
  \BibitemOpen
  \bibfield  {author} {\bibinfo {author} {\bibfnamefont {A.~R.}\ \bibnamefont
  {Hastie}}, \bibinfo {author} {\bibfnamefont {L.}~\bibnamefont {Dong}},
  \bibinfo {author} {\bibfnamefont {A.}~\bibnamefont {Smith}}, \bibinfo
  {author} {\bibfnamefont {J.}~\bibnamefont {Finklestein}}, \bibinfo {author}
  {\bibfnamefont {E.~T.}\ \bibnamefont {Lam}}, \bibinfo {author} {\bibfnamefont
  {N.}~\bibnamefont {Huo}}, \bibinfo {author} {\bibfnamefont {H.}~\bibnamefont
  {Cao}}, \bibinfo {author} {\bibfnamefont {P.-Y.}\ \bibnamefont {Kwok}},
  \bibinfo {author} {\bibfnamefont {K.~R.}\ \bibnamefont {Deal}}, \ and\
  \bibinfo {author} {\bibfnamefont {J.}~\bibnamefont {Dvorak}},\ }\href@noop {}
  {\bibfield  {journal} {\bibinfo  {journal} {PloS one}\ }\textbf {\bibinfo
  {volume} {8}},\ \bibinfo {pages} {e55864} (\bibinfo {year}
  {2013})}\BibitemShut {NoStop}%
\bibitem [{\citenamefont {Dorfman}(2013)}]{dorfman2013fluid}%
  \BibitemOpen
  \bibfield  {author} {\bibinfo {author} {\bibfnamefont {K.~D.}\ \bibnamefont
  {Dorfman}},\ }\href@noop {} {\bibfield  {journal} {\bibinfo  {journal} {AIChE
  J.}\ }\textbf {\bibinfo {volume} {59}},\ \bibinfo {pages} {346} (\bibinfo
  {year} {2013})}\BibitemShut {NoStop}%
\bibitem [{\citenamefont {M{\"u}ller}\ and\ \citenamefont
  {Westerlund}(2017)}]{muller2017optical}%
  \BibitemOpen
  \bibfield  {author} {\bibinfo {author} {\bibfnamefont {V.}~\bibnamefont
  {M{\"u}ller}}\ and\ \bibinfo {author} {\bibfnamefont {F.}~\bibnamefont
  {Westerlund}},\ }\href {\doibase 10.1039/C6LC01439A} {\bibfield  {journal}
  {\bibinfo  {journal} {Lab Chip}\ }\textbf {\bibinfo {volume} {17}},\ \bibinfo
  {pages} {579} (\bibinfo {year} {2017})}\BibitemShut {NoStop}%
\bibitem [{\citenamefont {Jun}\ and\ \citenamefont
  {Wright}(2010)}]{jun2010entropy}%
  \BibitemOpen
  \bibfield  {author} {\bibinfo {author} {\bibfnamefont {S.}~\bibnamefont
  {Jun}}\ and\ \bibinfo {author} {\bibfnamefont {A.}~\bibnamefont {Wright}},\
  }\href@noop {} {\bibfield  {journal} {\bibinfo  {journal} {Nat. Rev.
  Microbiol.}\ }\textbf {\bibinfo {volume} {8}},\ \bibinfo {pages} {600}
  (\bibinfo {year} {2010})}\BibitemShut {NoStop}%
\bibitem [{\citenamefont {Di~Ventura}\ \emph {et~al.}(2013)\citenamefont
  {Di~Ventura}, \citenamefont {Knecht}, \citenamefont {Andreas}, \citenamefont
  {Godinez}, \citenamefont {Fritsche}, \citenamefont {Rohr}, \citenamefont
  {Nickel}, \citenamefont {Heermann},\ and\ \citenamefont
  {Sourjik}}]{diventura2013chromosome}%
  \BibitemOpen
  \bibfield  {author} {\bibinfo {author} {\bibfnamefont {B.}~\bibnamefont
  {Di~Ventura}}, \bibinfo {author} {\bibfnamefont {B.}~\bibnamefont {Knecht}},
  \bibinfo {author} {\bibfnamefont {H.}~\bibnamefont {Andreas}}, \bibinfo
  {author} {\bibfnamefont {W.~J.}\ \bibnamefont {Godinez}}, \bibinfo {author}
  {\bibfnamefont {M.}~\bibnamefont {Fritsche}}, \bibinfo {author}
  {\bibfnamefont {K.}~\bibnamefont {Rohr}}, \bibinfo {author} {\bibfnamefont
  {W.}~\bibnamefont {Nickel}}, \bibinfo {author} {\bibfnamefont {D.~W.}\
  \bibnamefont {Heermann}}, \ and\ \bibinfo {author} {\bibfnamefont
  {V.}~\bibnamefont {Sourjik}},\ }\href@noop {} {\bibfield  {journal} {\bibinfo
   {journal} {Mol. Syst. Biol.}\ }\textbf {\bibinfo {volume} {9}},\ \bibinfo
  {pages} {686} (\bibinfo {year} {2013})}\BibitemShut {NoStop}%
\bibitem [{\citenamefont {Youngren}\ \emph {et~al.}(2014)\citenamefont
  {Youngren}, \citenamefont {Nielsen}, \citenamefont {Jun},\ and\ \citenamefont
  {Austin}}]{youngren2014multifork}%
  \BibitemOpen
  \bibfield  {author} {\bibinfo {author} {\bibfnamefont {B.}~\bibnamefont
  {Youngren}}, \bibinfo {author} {\bibfnamefont {H.~J.}\ \bibnamefont
  {Nielsen}}, \bibinfo {author} {\bibfnamefont {S.}~\bibnamefont {Jun}}, \ and\
  \bibinfo {author} {\bibfnamefont {S.}~\bibnamefont {Austin}},\ }\href@noop {}
  {\bibfield  {journal} {\bibinfo  {journal} {Genes Dev.}\ }\textbf {\bibinfo
  {volume} {28}},\ \bibinfo {pages} {71} (\bibinfo {year} {2014})}\BibitemShut
  {NoStop}%
\bibitem [{\citenamefont {M{a}nnik}\ \emph {et~al.}(2016)\citenamefont
  {M{a}nnik}, \citenamefont {Castillo}, \citenamefont {Yang}, \citenamefont
  {Siopsis},\ and\ \citenamefont {M{\"a}nnik}}]{mannik2016role}%
  \BibitemOpen
  \bibfield  {author} {\bibinfo {author} {\bibfnamefont {J.}~\bibnamefont
  {M{a}nnik}}, \bibinfo {author} {\bibfnamefont {D.~E.}\ \bibnamefont
  {Castillo}}, \bibinfo {author} {\bibfnamefont {D.}~\bibnamefont {Yang}},
  \bibinfo {author} {\bibfnamefont {G.}~\bibnamefont {Siopsis}}, \ and\
  \bibinfo {author} {\bibfnamefont {J.}~\bibnamefont {M{\"a}nnik}},\
  }\href@noop {} {\bibfield  {journal} {\bibinfo  {journal} {Nucleic Acids
  Res.}\ }\textbf {\bibinfo {volume} {44}},\ \bibinfo {pages} {1216} (\bibinfo
  {year} {2016})}\BibitemShut {NoStop}%
\bibitem [{\citenamefont {Cass}\ \emph {et~al.}(2016)\citenamefont {Cass},
  \citenamefont {Kuwada}, \citenamefont {Traxler},\ and\ \citenamefont
  {Wiggins}}]{cass2016escherichia}%
  \BibitemOpen
  \bibfield  {author} {\bibinfo {author} {\bibfnamefont {J.~A.}\ \bibnamefont
  {Cass}}, \bibinfo {author} {\bibfnamefont {N.~J.}\ \bibnamefont {Kuwada}},
  \bibinfo {author} {\bibfnamefont {B.}~\bibnamefont {Traxler}}, \ and\
  \bibinfo {author} {\bibfnamefont {P.~A.}\ \bibnamefont {Wiggins}},\
  }\href@noop {} {\bibfield  {journal} {\bibinfo  {journal} {Biophys. J.}\
  }\textbf {\bibinfo {volume} {110}},\ \bibinfo {pages} {2597} (\bibinfo {year}
  {2016})}\BibitemShut {NoStop}%
\bibitem [{\citenamefont {Wu}\ \emph {et~al.}(2019)\citenamefont {Wu},
  \citenamefont {Swain}, \citenamefont {Kuijpers}, \citenamefont {Zheng},
  \citenamefont {Felter}, \citenamefont {Guurink}, \citenamefont {Solari},
  \citenamefont {Jun}, \citenamefont {Shimizu}, \citenamefont {Chaudhuri} \emph
  {et~al.}}]{wu2019cell}%
  \BibitemOpen
  \bibfield  {author} {\bibinfo {author} {\bibfnamefont {F.}~\bibnamefont
  {Wu}}, \bibinfo {author} {\bibfnamefont {P.}~\bibnamefont {Swain}}, \bibinfo
  {author} {\bibfnamefont {L.}~\bibnamefont {Kuijpers}}, \bibinfo {author}
  {\bibfnamefont {X.}~\bibnamefont {Zheng}}, \bibinfo {author} {\bibfnamefont
  {K.}~\bibnamefont {Felter}}, \bibinfo {author} {\bibfnamefont
  {M.}~\bibnamefont {Guurink}}, \bibinfo {author} {\bibfnamefont
  {J.}~\bibnamefont {Solari}}, \bibinfo {author} {\bibfnamefont
  {S.}~\bibnamefont {Jun}}, \bibinfo {author} {\bibfnamefont {T.~S.}\
  \bibnamefont {Shimizu}}, \bibinfo {author} {\bibfnamefont {D.}~\bibnamefont
  {Chaudhuri}},  \emph {et~al.},\ }\href@noop {} {\bibfield  {journal}
  {\bibinfo  {journal} {Curr. Biol.}\ }\textbf {\bibinfo {volume} {29}},\
  \bibinfo {pages} {2131} (\bibinfo {year} {2019})}\BibitemShut {NoStop}%
\bibitem [{\citenamefont {Wu}\ \emph {et~al.}(2020)\citenamefont {Wu},
  \citenamefont {Lee}, \citenamefont {Park}, \citenamefont {Eland},
  \citenamefont {Wipat}, \citenamefont {Holden},\ and\ \citenamefont
  {Errington}}]{wu2020geometric}%
  \BibitemOpen
  \bibfield  {author} {\bibinfo {author} {\bibfnamefont {L.~J.}\ \bibnamefont
  {Wu}}, \bibinfo {author} {\bibfnamefont {S.}~\bibnamefont {Lee}}, \bibinfo
  {author} {\bibfnamefont {S.}~\bibnamefont {Park}}, \bibinfo {author}
  {\bibfnamefont {L.}~\bibnamefont {Eland}}, \bibinfo {author} {\bibfnamefont
  {A.}~\bibnamefont {Wipat}}, \bibinfo {author} {\bibfnamefont
  {S.}~\bibnamefont {Holden}}, \ and\ \bibinfo {author} {\bibfnamefont
  {J.}~\bibnamefont {Errington}},\ }\href@noop {} {\bibfield  {journal}
  {\bibinfo  {journal} {Nat. Commun.}\ }\textbf {\bibinfo {volume} {11}},\
  \bibinfo {pages} {1} (\bibinfo {year} {2020})}\BibitemShut {NoStop}%
\bibitem [{\citenamefont {El~Najjar}\ \emph {et~al.}(2020)\citenamefont
  {El~Najjar}, \citenamefont {Geisel}, \citenamefont {Schmidt}, \citenamefont
  {Dersch}, \citenamefont {Mayer}, \citenamefont {Hartmann}, \citenamefont
  {Eckhardt}, \citenamefont {Lenz},\ and\ \citenamefont
  {Graumann}}]{elnajjar2020chromosome}%
  \BibitemOpen
  \bibfield  {author} {\bibinfo {author} {\bibfnamefont {N.}~\bibnamefont
  {El~Najjar}}, \bibinfo {author} {\bibfnamefont {D.}~\bibnamefont {Geisel}},
  \bibinfo {author} {\bibfnamefont {F.}~\bibnamefont {Schmidt}}, \bibinfo
  {author} {\bibfnamefont {S.}~\bibnamefont {Dersch}}, \bibinfo {author}
  {\bibfnamefont {B.}~\bibnamefont {Mayer}}, \bibinfo {author} {\bibfnamefont
  {R.}~\bibnamefont {Hartmann}}, \bibinfo {author} {\bibfnamefont
  {B.}~\bibnamefont {Eckhardt}}, \bibinfo {author} {\bibfnamefont
  {P.}~\bibnamefont {Lenz}}, \ and\ \bibinfo {author} {\bibfnamefont {P.~L.}\
  \bibnamefont {Graumann}},\ }\href@noop {} {\bibfield  {journal} {\bibinfo
  {journal} {mSphere}\ }\textbf {\bibinfo {volume} {5}},\ \bibinfo {pages}
  {e00255} (\bibinfo {year} {2020})}\BibitemShut {NoStop}%
\bibitem [{\citenamefont {Japaridze}\ \emph {et~al.}(2020)\citenamefont
  {Japaridze}, \citenamefont {Gogou}, \citenamefont {Kerssemakers},
  \citenamefont {Nguyen},\ and\ \citenamefont {Dekker}}]{japaridze2020direct}%
  \BibitemOpen
  \bibfield  {author} {\bibinfo {author} {\bibfnamefont {A.}~\bibnamefont
  {Japaridze}}, \bibinfo {author} {\bibfnamefont {C.}~\bibnamefont {Gogou}},
  \bibinfo {author} {\bibfnamefont {J.}~\bibnamefont {Kerssemakers}}, \bibinfo
  {author} {\bibfnamefont {H.~M.}\ \bibnamefont {Nguyen}}, \ and\ \bibinfo
  {author} {\bibfnamefont {C.}~\bibnamefont {Dekker}},\ }\href@noop {}
  {\bibfield  {journal} {\bibinfo  {journal} {Nat. Commun.}\ }\textbf {\bibinfo
  {volume} {11}},\ \bibinfo {pages} {1} (\bibinfo {year} {2020})}\BibitemShut
  {NoStop}%
\bibitem [{\citenamefont {Liang}\ \emph {et~al.}(2020)\citenamefont {Liang},
  \citenamefont {Quan}, \citenamefont {Li}, \citenamefont {Loton},
  \citenamefont {Bredeche}, \citenamefont {Lindner},\ and\ \citenamefont
  {Xu}}]{liang2020artificial}%
  \BibitemOpen
  \bibfield  {author} {\bibinfo {author} {\bibfnamefont {B.}~\bibnamefont
  {Liang}}, \bibinfo {author} {\bibfnamefont {B.}~\bibnamefont {Quan}},
  \bibinfo {author} {\bibfnamefont {J.}~\bibnamefont {Li}}, \bibinfo {author}
  {\bibfnamefont {C.}~\bibnamefont {Loton}}, \bibinfo {author} {\bibfnamefont
  {M.-F.}\ \bibnamefont {Bredeche}}, \bibinfo {author} {\bibfnamefont {A.~B.}\
  \bibnamefont {Lindner}}, \ and\ \bibinfo {author} {\bibfnamefont
  {L.}~\bibnamefont {Xu}},\ }\href@noop {} {\bibfield  {journal} {\bibinfo
  {journal} {Sci. Rep.}\ }\textbf {\bibinfo {volume} {10}},\ \bibinfo {pages}
  {1} (\bibinfo {year} {2020})}\BibitemShut {NoStop}%
\bibitem [{\citenamefont {Gogou}\ \emph {et~al.}(2021)\citenamefont {Gogou},
  \citenamefont {Japaridze},\ and\ \citenamefont
  {Dekker}}]{gogou2021mechanisms}%
  \BibitemOpen
  \bibfield  {author} {\bibinfo {author} {\bibfnamefont {C.}~\bibnamefont
  {Gogou}}, \bibinfo {author} {\bibfnamefont {A.}~\bibnamefont {Japaridze}}, \
  and\ \bibinfo {author} {\bibfnamefont {C.}~\bibnamefont {Dekker}},\
  }\href@noop {} {\bibfield  {journal} {\bibinfo  {journal} {Front.
  Microbiol.}\ }\textbf {\bibinfo {volume} {12}},\ \bibinfo {pages} {1533}
  (\bibinfo {year} {2021})}\BibitemShut {NoStop}%
\bibitem [{\citenamefont {Klotz}\ \emph
  {et~al.}(2015{\natexlab{a}})\citenamefont {Klotz}, \citenamefont {Mamaev},
  \citenamefont {Duong}, \citenamefont {de~Haan},\ and\ \citenamefont
  {Reisner}}]{klotz2015correlated}%
  \BibitemOpen
  \bibfield  {author} {\bibinfo {author} {\bibfnamefont {A.~R.}\ \bibnamefont
  {Klotz}}, \bibinfo {author} {\bibfnamefont {M.}~\bibnamefont {Mamaev}},
  \bibinfo {author} {\bibfnamefont {L.}~\bibnamefont {Duong}}, \bibinfo
  {author} {\bibfnamefont {H.~W.}\ \bibnamefont {de~Haan}}, \ and\ \bibinfo
  {author} {\bibfnamefont {W.~W.}\ \bibnamefont {Reisner}},\ }\href@noop {}
  {\bibfield  {journal} {\bibinfo  {journal} {Macromolecules}\ }\textbf
  {\bibinfo {volume} {48}},\ \bibinfo {pages} {4742} (\bibinfo {year}
  {2015}{\natexlab{a}})}\BibitemShut {NoStop}%
\bibitem [{\citenamefont {Klotz}\ \emph
  {et~al.}(2015{\natexlab{b}})\citenamefont {Klotz}, \citenamefont {Duong},
  \citenamefont {Mamaev}, \citenamefont {de~Haan}, \citenamefont {Chen},\ and\
  \citenamefont {Reisner}}]{klotz2015measuring}%
  \BibitemOpen
  \bibfield  {author} {\bibinfo {author} {\bibfnamefont {A.~R.}\ \bibnamefont
  {Klotz}}, \bibinfo {author} {\bibfnamefont {L.}~\bibnamefont {Duong}},
  \bibinfo {author} {\bibfnamefont {M.}~\bibnamefont {Mamaev}}, \bibinfo
  {author} {\bibfnamefont {H.~W.}\ \bibnamefont {de~Haan}}, \bibinfo {author}
  {\bibfnamefont {J.~Z.}\ \bibnamefont {Chen}}, \ and\ \bibinfo {author}
  {\bibfnamefont {W.~W.}\ \bibnamefont {Reisner}},\ }\href@noop {} {\bibfield
  {journal} {\bibinfo  {journal} {Macromolecules}\ }\textbf {\bibinfo {volume}
  {48}},\ \bibinfo {pages} {5028} (\bibinfo {year}
  {2015}{\natexlab{b}})}\BibitemShut {NoStop}%
\bibitem [{\citenamefont {Berard}\ \emph {et~al.}(2014)\citenamefont {Berard},
  \citenamefont {Michaud}, \citenamefont {Mahshid}, \citenamefont {Ahamed},
  \citenamefont {McFaul}, \citenamefont {Leith}, \citenamefont
  {B{\'e}rub{\'e}}, \citenamefont {Sladek}, \citenamefont {Reisner},\ and\
  \citenamefont {Leslie}}]{berard2014convex}%
  \BibitemOpen
  \bibfield  {author} {\bibinfo {author} {\bibfnamefont {D.~J.}\ \bibnamefont
  {Berard}}, \bibinfo {author} {\bibfnamefont {F.}~\bibnamefont {Michaud}},
  \bibinfo {author} {\bibfnamefont {S.}~\bibnamefont {Mahshid}}, \bibinfo
  {author} {\bibfnamefont {M.~J.}\ \bibnamefont {Ahamed}}, \bibinfo {author}
  {\bibfnamefont {C.~M.}\ \bibnamefont {McFaul}}, \bibinfo {author}
  {\bibfnamefont {J.~S.}\ \bibnamefont {Leith}}, \bibinfo {author}
  {\bibfnamefont {P.}~\bibnamefont {B{\'e}rub{\'e}}}, \bibinfo {author}
  {\bibfnamefont {R.}~\bibnamefont {Sladek}}, \bibinfo {author} {\bibfnamefont
  {W.}~\bibnamefont {Reisner}}, \ and\ \bibinfo {author} {\bibfnamefont
  {S.~R.}\ \bibnamefont {Leslie}},\ }\href@noop {} {\bibfield  {journal}
  {\bibinfo  {journal} {Proc. Natl. Acad. Sci. U.S.A.}\ }\textbf {\bibinfo
  {volume} {111}},\ \bibinfo {pages} {13295} (\bibinfo {year}
  {2014})}\BibitemShut {NoStop}%
\bibitem [{\citenamefont {Capaldi}\ \emph {et~al.}(2018)\citenamefont
  {Capaldi}, \citenamefont {Liu}, \citenamefont {Zhang}, \citenamefont {Zeng},
  \citenamefont {Reyes-Lamothe},\ and\ \citenamefont
  {Reisner}}]{capaldi2018probing}%
  \BibitemOpen
  \bibfield  {author} {\bibinfo {author} {\bibfnamefont {X.}~\bibnamefont
  {Capaldi}}, \bibinfo {author} {\bibfnamefont {Z.}~\bibnamefont {Liu}},
  \bibinfo {author} {\bibfnamefont {Y.}~\bibnamefont {Zhang}}, \bibinfo
  {author} {\bibfnamefont {L.}~\bibnamefont {Zeng}}, \bibinfo {author}
  {\bibfnamefont {R.}~\bibnamefont {Reyes-Lamothe}}, \ and\ \bibinfo {author}
  {\bibfnamefont {W.}~\bibnamefont {Reisner}},\ }\href@noop {} {\bibfield
  {journal} {\bibinfo  {journal} {{Soft Matter}}\ }\textbf {\bibinfo {volume}
  {14}},\ \bibinfo {pages} {8455} (\bibinfo {year} {2018})}\BibitemShut
  {NoStop}%
\bibitem [{\citenamefont {Liu}\ \emph {et~al.}(2022)\citenamefont {Liu},
  \citenamefont {Capaldi}, \citenamefont {Zeng}, \citenamefont {Zhang},
  \citenamefont {Reyes-Lamothe},\ and\ \citenamefont
  {Reisner}}]{liu2022confinement}%
  \BibitemOpen
  \bibfield  {author} {\bibinfo {author} {\bibfnamefont {Z.}~\bibnamefont
  {Liu}}, \bibinfo {author} {\bibfnamefont {X.}~\bibnamefont {Capaldi}},
  \bibinfo {author} {\bibfnamefont {L.}~\bibnamefont {Zeng}}, \bibinfo {author}
  {\bibfnamefont {Y.}~\bibnamefont {Zhang}}, \bibinfo {author} {\bibfnamefont
  {R.}~\bibnamefont {Reyes-Lamothe}}, \ and\ \bibinfo {author} {\bibfnamefont
  {W.}~\bibnamefont {Reisner}},\ }\href@noop {} {\bibfield  {journal} {\bibinfo
   {journal} {Nat. Commun.}\ }\textbf {\bibinfo {volume} {13}},\ \bibinfo
  {pages} {1} (\bibinfo {year} {2022})}\BibitemShut {NoStop}%
\bibitem [{\citenamefont {Polson}\ and\ \citenamefont
  {Rehel}(2021)}]{polson2021equilibrium}%
  \BibitemOpen
  \bibfield  {author} {\bibinfo {author} {\bibfnamefont {J.~M.}\ \bibnamefont
  {Polson}}\ and\ \bibinfo {author} {\bibfnamefont {D.~A.}\ \bibnamefont
  {Rehel}},\ }\href@noop {} {\bibfield  {journal} {\bibinfo  {journal} {Soft
  Matter}\ }\textbf {\bibinfo {volume} {17}},\ \bibinfo {pages} {5792}
  (\bibinfo {year} {2021})}\BibitemShut {NoStop}%
\bibitem [{\citenamefont {Micheletti}\ \emph {et~al.}(2011)\citenamefont
  {Micheletti}, \citenamefont {Marenduzzo},\ and\ \citenamefont
  {Orlandini}}]{micheletti2011polymers}%
  \BibitemOpen
  \bibfield  {author} {\bibinfo {author} {\bibfnamefont {C.}~\bibnamefont
  {Micheletti}}, \bibinfo {author} {\bibfnamefont {D.}~\bibnamefont
  {Marenduzzo}}, \ and\ \bibinfo {author} {\bibfnamefont {E.}~\bibnamefont
  {Orlandini}},\ }\href@noop {} {\bibfield  {journal} {\bibinfo  {journal}
  {Phys. Rep.}\ }\textbf {\bibinfo {volume} {504}},\ \bibinfo {pages} {1}
  (\bibinfo {year} {2011})}\BibitemShut {NoStop}%
\bibitem [{\citenamefont {Tree}\ \emph {et~al.}(2013)\citenamefont {Tree},
  \citenamefont {Muralidhar}, \citenamefont {Doyle},\ and\ \citenamefont
  {Dorfman}}]{tree2013dna}%
  \BibitemOpen
  \bibfield  {author} {\bibinfo {author} {\bibfnamefont {D.~R.}\ \bibnamefont
  {Tree}}, \bibinfo {author} {\bibfnamefont {A.}~\bibnamefont {Muralidhar}},
  \bibinfo {author} {\bibfnamefont {P.~S.}\ \bibnamefont {Doyle}}, \ and\
  \bibinfo {author} {\bibfnamefont {K.~D.}\ \bibnamefont {Dorfman}},\
  }\href@noop {} {\bibfield  {journal} {\bibinfo  {journal} {Macromolecules}\
  }\textbf {\bibinfo {volume} {46}},\ \bibinfo {pages} {8369} (\bibinfo {year}
  {2013})}\BibitemShut {NoStop}%
\bibitem [{\citenamefont {Dobrynin}(2006)}]{dobrynin2006effect}%
  \BibitemOpen
  \bibfield  {author} {\bibinfo {author} {\bibfnamefont {A.~V.}\ \bibnamefont
  {Dobrynin}},\ }\href@noop {} {\bibfield  {journal} {\bibinfo  {journal}
  {Macromolecules}\ }\textbf {\bibinfo {volume} {39}},\ \bibinfo {pages} {9519}
  (\bibinfo {year} {2006})}\BibitemShut {NoStop}%
\bibitem [{\citenamefont {Kundukad}\ \emph {et~al.}(2014)\citenamefont
  {Kundukad}, \citenamefont {Yan},\ and\ \citenamefont
  {Doyle}}]{kundukad2014effect}%
  \BibitemOpen
  \bibfield  {author} {\bibinfo {author} {\bibfnamefont {B.}~\bibnamefont
  {Kundukad}}, \bibinfo {author} {\bibfnamefont {J.}~\bibnamefont {Yan}}, \
  and\ \bibinfo {author} {\bibfnamefont {P.~S.}\ \bibnamefont {Doyle}},\
  }\href@noop {} {\bibfield  {journal} {\bibinfo  {journal} {Soft matter}\
  }\textbf {\bibinfo {volume} {10}},\ \bibinfo {pages} {9721} (\bibinfo {year}
  {2014})}\BibitemShut {NoStop}%
\bibitem [{\citenamefont {Wang}\ \emph {et~al.}(1998)\citenamefont {Wang},
  \citenamefont {Lin},\ and\ \citenamefont {Schwartz}}]{wang1998scanning}%
  \BibitemOpen
  \bibfield  {author} {\bibinfo {author} {\bibfnamefont {W.}~\bibnamefont
  {Wang}}, \bibinfo {author} {\bibfnamefont {J.}~\bibnamefont {Lin}}, \ and\
  \bibinfo {author} {\bibfnamefont {D.~C.}\ \bibnamefont {Schwartz}},\
  }\href@noop {} {\bibfield  {journal} {\bibinfo  {journal} {Biophys. J.}\
  }\textbf {\bibinfo {volume} {75}},\ \bibinfo {pages} {513} (\bibinfo {year}
  {1998})}\BibitemShut {NoStop}%
\bibitem [{\citenamefont {Frenkel}\ and\ \citenamefont
  {Smit}(2002)}]{frenkel2002understanding}%
  \BibitemOpen
  \bibfield  {author} {\bibinfo {author} {\bibfnamefont {D.}~\bibnamefont
  {Frenkel}}\ and\ \bibinfo {author} {\bibfnamefont {B.}~\bibnamefont {Smit}},\
  }\href@noop {} {\emph {\bibinfo {title} {Understanding Molecular Simulation:
  From Algorithms to Applications}}},\ \bibinfo {edition} {2nd}\ ed.\ (\bibinfo
   {publisher} {Academic Press},\ \bibinfo {address} {London},\ \bibinfo {year}
  {2002})\ Chap.~\bibinfo {chapter} {7}\BibitemShut {NoStop}%
\bibitem [{\citenamefont {Polson}\ and\ \citenamefont
  {McLure}(2019)}]{polson2019free}%
  \BibitemOpen
  \bibfield  {author} {\bibinfo {author} {\bibfnamefont {J.~M.}\ \bibnamefont
  {Polson}}\ and\ \bibinfo {author} {\bibfnamefont {Z.~R.}\ \bibnamefont
  {McLure}},\ }\href@noop {} {\bibfield  {journal} {\bibinfo  {journal} {Phys.
  Rev. E}\ }\textbf {\bibinfo {volume} {99}},\ \bibinfo {pages} {062503}
  (\bibinfo {year} {2019})}\BibitemShut {NoStop}%
\bibitem [{\citenamefont {Phillips}\ \emph {et~al.}(2012)\citenamefont
  {Phillips}, \citenamefont {Theriot}, \citenamefont {Garcia},\ and\
  \citenamefont {Kondev}}]{Phillips_book}%
  \BibitemOpen
  \bibfield  {author} {\bibinfo {author} {\bibfnamefont {R.}~\bibnamefont
  {Phillips}}, \bibinfo {author} {\bibfnamefont {J.}~\bibnamefont {Theriot}},
  \bibinfo {author} {\bibfnamefont {J.~H.}\ \bibnamefont {Garcia}}, \ and\
  \bibinfo {author} {\bibfnamefont {J.}~\bibnamefont {Kondev}},\ }\href@noop {}
  {\emph {\bibinfo {title} {Physical Biology of the Cell}}},\ \bibinfo
  {edition} {2nd}\ ed.\ (\bibinfo  {publisher} {Garland Science},\ \bibinfo
  {address} {New York},\ \bibinfo {year} {2012})\BibitemShut {NoStop}%
\bibitem [{\citenamefont {Matkowsky}\ \emph {et~al.}(1988)\citenamefont
  {Matkowsky}, \citenamefont {Nitzan},\ and\ \citenamefont
  {Schuss}}]{matkowsky1988does}%
  \BibitemOpen
  \bibfield  {author} {\bibinfo {author} {\bibfnamefont {B.~J.}\ \bibnamefont
  {Matkowsky}}, \bibinfo {author} {\bibfnamefont {A.}~\bibnamefont {Nitzan}}, \
  and\ \bibinfo {author} {\bibfnamefont {Z.}~\bibnamefont {Schuss}},\
  }\href@noop {} {\bibfield  {journal} {\bibinfo  {journal} {J. Chem. Phys}\
  }\textbf {\bibinfo {volume} {88}},\ \bibinfo {pages} {4765} (\bibinfo {year}
  {1988})}\BibitemShut {NoStop}%
\bibitem [{\citenamefont {de~Gennes}(1979)}]{deGennes_book}%
  \BibitemOpen
  \bibfield  {author} {\bibinfo {author} {\bibfnamefont {P.}~\bibnamefont
  {de~Gennes}},\ }\href@noop {} {\emph {\bibinfo {title} {{Scaling Concepts in
  Polymer Physics}}}}\ (\bibinfo  {publisher} {Cornell University Press},\
  \bibinfo {address} {Ithica NY},\ \bibinfo {year} {1979})\BibitemShut
  {NoStop}%
\bibitem [{\citenamefont {Reyes-Lamothe}\ \emph {et~al.}(2013)\citenamefont
  {Reyes-Lamothe}, \citenamefont {Tran}, \citenamefont {Meas}, \citenamefont
  {Lee}, \citenamefont {Li}, \citenamefont {Sherratt},\ and\ \citenamefont
  {Tolmasky}}]{reyes2013high}%
  \BibitemOpen
  \bibfield  {author} {\bibinfo {author} {\bibfnamefont {R.}~\bibnamefont
  {Reyes-Lamothe}}, \bibinfo {author} {\bibfnamefont {T.}~\bibnamefont {Tran}},
  \bibinfo {author} {\bibfnamefont {D.}~\bibnamefont {Meas}}, \bibinfo {author}
  {\bibfnamefont {L.}~\bibnamefont {Lee}}, \bibinfo {author} {\bibfnamefont
  {A.~M.}\ \bibnamefont {Li}}, \bibinfo {author} {\bibfnamefont {D.~J.}\
  \bibnamefont {Sherratt}}, \ and\ \bibinfo {author} {\bibfnamefont {M.~E.}\
  \bibnamefont {Tolmasky}},\ }\href@noop {} {\bibfield  {journal} {\bibinfo
  {journal} {Nucleic Acids Res.}\ }\textbf {\bibinfo {volume} {42}},\ \bibinfo
  {pages} {1042} (\bibinfo {year} {2013})}\BibitemShut {NoStop}%
\end{thebibliography}%


\begin{thebibliography}{1}%
\makeatletter
\providecommand \@ifxundefined [1]{%
 \@ifx{#1\undefined}
}%
\providecommand \@ifnum [1]{%
 \ifnum #1\expandafter \@firstoftwo
 \else \expandafter \@secondoftwo
 \fi
}%
\providecommand \@ifx [1]{%
 \ifx #1\expandafter \@firstoftwo
 \else \expandafter \@secondoftwo
 \fi
}%
\providecommand \natexlab [1]{#1}%
\providecommand \enquote  [1]{``#1''}%
\providecommand \bibnamefont  [1]{#1}%
\providecommand \bibfnamefont [1]{#1}%
\providecommand \citenamefont [1]{#1}%
\providecommand \href@noop [0]{\@secondoftwo}%
\providecommand \href [0]{\begingroup \@sanitize@url \@href}%
\providecommand \@href[1]{\@@startlink{#1}\@@href}%
\providecommand \@@href[1]{\endgroup#1\@@endlink}%
\providecommand \@sanitize@url [0]{\catcode `\\12\catcode `\$12\catcode
  `\&12\catcode `\#12\catcode `\^12\catcode `\_12\catcode `\%12\relax}%
\providecommand \@@startlink[1]{}%
\providecommand \@@endlink[0]{}%
\providecommand \url  [0]{\begingroup\@sanitize@url \@url }%
\providecommand \@url [1]{\endgroup\@href {#1}{\urlprefix }}%
\providecommand \urlprefix  [0]{URL }%
\providecommand \Eprint [0]{\href }%
\providecommand \doibase [0]{http://dx.doi.org/}%
\providecommand \selectlanguage [0]{\@gobble}%
\providecommand \bibinfo  [0]{\@secondoftwo}%
\providecommand \bibfield  [0]{\@secondoftwo}%
\providecommand \translation [1]{[#1]}%
\providecommand \BibitemOpen [0]{}%
\providecommand \bibitemStop [0]{}%
\providecommand \bibitemNoStop [0]{.\EOS\space}%
\providecommand \EOS [0]{\spacefactor3000\relax}%
\providecommand \BibitemShut  [1]{\csname bibitem#1\endcsname}%
\let\auto@bib@innerbib\@empty
\bibitem [{\citenamefont {Matkowsky}\ \emph {et~al.}(1988)\citenamefont
  {Matkowsky}, \citenamefont {Nitzan},\ and\ \citenamefont
  {Schuss}}]{matkowsky1988does}%
  \BibitemOpen
  \bibfield  {author} {\bibinfo {author} {\bibfnamefont {B.~J.}\ \bibnamefont
  {Matkowsky}}, \bibinfo {author} {\bibfnamefont {A.}~\bibnamefont {Nitzan}}, \
  and\ \bibinfo {author} {\bibfnamefont {Z.}~\bibnamefont {Schuss}},\
  }\href@noop {} {\bibfield  {journal} {\bibinfo  {journal} {J. Chem. Phys}\
  }\textbf {\bibinfo {volume} {88}},\ \bibinfo {pages} {4765} (\bibinfo {year}
  {1988})}\BibitemShut {NoStop}%
\end{thebibliography}%

%

\end{document}


\title{Supplemental Material for: ``Equilibrium behaviour of two cavity-confined polymers:
Effects of polymer width and system asymmetries''}

\author{Desiree Rehel}
\affiliation{Department of Physics, University of Prince Edward Island,
550 University Ave., Charlottetown, Prince Edward Island, C1A 4P3, Canada}
\author{James M. Polson}
\affiliation{Department of Physics, University of Prince Edward Island,
550 University Ave., Charlottetown, Prince Edward Island, C1A 4P3, Canada}


\maketitle

\section{Clarification of the calculation employing a 2D generalization of Kramers theory}

In Section~4.3 of the article, we examined the dynamics of two polymers confined to 
a cavity with a rectangular cross-section. We found that the polymers swap positions
on either end of the box with a dwell-time distribution that is characterized by a time
constant, $\tau_{\rm d}$. The time constant varies with the changes in the dimensions
of the cavity as a result in concomitant changes in the underlying free-energy landscape.
To understand the relationship between $\tau_{\rm d}$ and the free energy, we 
employed a multidimensional generalization of Kramers theory.\cite{matkowsky1988does}
Here, we clarify some details of the calculation used in that analysis.

As noted in Section~4.3 of the article, the theory predicts that
\begin{eqnarray}
e^{\beta\Delta F} = D\tau_\text{d}^*,
\label{eq:Kramer-Langer}
\end{eqnarray}
where
\begin{eqnarray}
\tau_{\rm d}^* \equiv \frac{Q_\text{B} \omega_\text{B} \omega_\text{W}}{2\pi Q_\text{W}} \tau_{\rm d}.
\end{eqnarray}
Here, $D$ is the Rouse diffusion coefficient, $\Delta F$ is the free energy barrier height,
and $\omega_\text{B}$ and $\omega_\text{W}$ represent the effective frequencies of the
well and barrier, respectively.  In addition, $Q_\text{B}$ and $Q_\text{W}$
are the partition functions associated with the non-reactive modes at the free energy
barrier and well, respectively.  The quantities $\omega_\text{W}$ and $\omega_\text{B}$
are obtained from the free energy function, $F(x,y)/k_{\rm B}T\equiv -\ln {\cal P}(x,y)$,
as follows. First, a cross section of the free energy function in the $y$
direction at $x$=0 is fit to a function of the form
$F(y)=A+B(y-y_{\rm min})^2+C(y-y_{\rm min})^3+D(y-y_{\rm min})^4$, where $y_{\rm min}$
is the position of the free energy minimum. From this fit, we choose $B=\omega_\text{W}^2/2$.
Next, we fit a cross section of the free energy along the minimum free energy path
in the vicinity of the saddle point at $y=0$ to the function $F(y)=A+By^2+Dy^4$, from
which we choose $-B=\omega_\text{B}^2/2$.  The quantity $Q_{\rm W}$ is obtained via
a numerical approximation to the the following integral over the nonreactive mode:
$\int_{-L_x/2}^{L_x/2} dx \exp[-(F(x,y_{\rm min})-F_{\rm min}))/k_{\rm B}T]$, where
$F_{\rm min}$ is the free energy at the minimum $(x=0,y=y_{\rm min})$. Likewise, the partition
function for the nonreactive mode at the barrier, $Q_{\rm B}$, located at $y=0$, is given by
$\int_{-L_x/2}^{L_x/2} dx \exp[-(F(x,0)-F_{\rm bar}))/k_{\rm B}T]$, where $F_{\rm bar}$
is the free energy at the barrier. Note that the free energy barrier height is of course
$\Delta F=F_{\rm bar}-F_{\rm min}$.

\section{Effect of topological asymmetry for polymers confined to an elliptical cavity}

In Section~4.4 of the article, we examined the behaviour of a long linear semiflexible polymer
and a short ring polymer, both confined to a cavity with an elliptical cross section. The linear
polymer had a length of $N_1=1000$ monomers and a bending rigidity of $\kappa=6.36$, and the
ring polymer had a length of $N_2=25$~monomers. In addition, the elliptical cross section 
of the cavity had an area of $A=3000$ and a height of $h=15$. (All quantities are given in the
reduced units defined in the article.) Various values of the eccentricity of the ellipse were
considered. 

\begin{figure*}[!ht]
\begin{center}
\vspace*{0.2in}
\includegraphics[angle=0,width=0.99\textwidth]{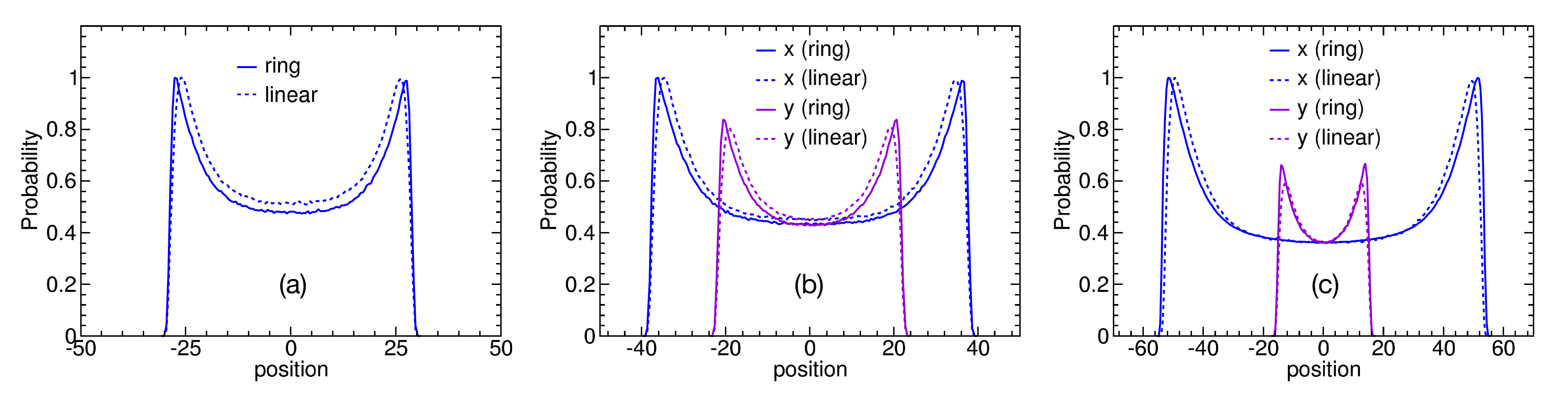}
\end{center}
\caption{Cross sections along the $x$ and $y$ axes (defined in the article) of the 
centre-of-mass probability distributions for a short polymer of length $N_2=25$ trapped with 
a long, linear, semiflexible polymer of length $N_1=1000$ and bending rigidity $\kappa=6.36$ 
in a cavity.  The cavity has an elliptical cross section of area $A=3000$ and height 
$h=15$.  Results are shown for eccentricities of (a) $e=0$, (b) $e=0.8$, and (c) 
$e=0.95$.  Each graph shows results for cases where the short polymer has linear 
and ring topologies. For convenience, distributions have been scaled so that the maximum 
of the underlying 2D distribution for ${\cal P}(x,y)$ is equal to one. Note the cross
sections for $x$ and $y$ are identical for $e=0$.}
\label{fig:figsupp}
\end{figure*}

The distribution of the ring polymer is largely determined by the fact that it is so much
shorter than the linear polymer.  Here, we examine the effects of the topological difference
between the two polymers. To do this, we carry out the same calculations that were described
in Section~4.4 with the single change that the topology short polymer is linear rather than
ring type.  Figure~\ref{fig:figsupp} shows cross sections of the 2D probability distributions
for the short polymer along the $x$ and $y$ axes (defined in the article) for cavity eccentricities
of $e=0, 0.8$ and $0.95$. The results clearly show that the topology of the short polymer has a
minimal effect on the distributions. The single notable effect is the slightly greater
degree of repulsion from the lateral walls for the linear polymer. This is evident in the
fact that the probability peaks are displaced slightly inward for the linear polymer relative
to that for the ring polymer.  This feature most likely arises because of the slightly
larger average size of the linear polymer.

\bibliography{paper}
\bibliographystyle{apsrev4-1}